\documentclass{emulateapj}

\slugcomment{Published in ApJ, 2008, 675, 1025}

\shorttitle{Evolution of AGN host galaxies}
\shortauthors{Silverman et al.}

\begin{document}


\title{The Evolution of AGN Host Galaxies: From Blue to Red and
the Influence of Large-Scale Structures}


\author{
J.~D.~Silverman,\altaffilmark{1}
V.~Mainieri,\altaffilmark{1,2}
B.~D.~Lehmer,\altaffilmark{3}
D.~M.~Alexander,\altaffilmark{4}
F.~E.~Bauer,\altaffilmark{5}
J.~Bergeron,\altaffilmark{6,12}
W.~N.~Brandt,\altaffilmark{3}
R.~Gilli,\altaffilmark{7} 
G.~Hasinger,\altaffilmark{1}
D.~P.~Schneider,\altaffilmark{3}
P.~Tozzi,\altaffilmark{8}
C.~Vignali,\altaffilmark{9}
A.~M.~Koekemoer,\altaffilmark{10}
T.~Miyaji,\altaffilmark{11}
P.~Popesso,\altaffilmark{2,12} 
P.~Rosati,\altaffilmark{2,12}
\& G.~Szokoly\altaffilmark{1}
}


\altaffiltext{1}{Max-Planck-Institut f\"ur extraterrestrische Physik,
D-84571 Garching, Germany} \altaffiltext{2}{European Southern
Observatory, Karl-Schwarzschild-Strasse 2, Garching, D-85748, Germany}
\altaffiltext{3}{Department of Astronomy \& Astrophysics, 525 Davey
Lab, The Pennsylvania State University, University Park, PA 16802,
USA} \altaffiltext{4}{Department of Physics, University of Durham,
South Road, Durham, DH1 3LE, UK} \altaffiltext{5}{Columbia
Astrophysics Laboratory, Columbia University, Pupin Labortories, 550
W. 120th St., Rm 1418, New York, NY 10027, USA}
\altaffiltext{6}{Institut d'Astrophysique de Paris, 98bis Boulevard,
F-75014 Paris, France} \altaffiltext{7}{Istituto Nazionale di
Astrofisica (INAF) - Osservatorio Astronomico di Bologna, Via Ranzani
1, 40127 Bologna, Italy} \altaffiltext{8}{INAF - Osservatorio
Astronomico di Trieste, via G. B. Tiepolo 11, 34131 Trieste, Italy}
\altaffiltext{9}{Dipartimento di Astronomia, Universit\'a degli Studi
di Bologna, Via Ranzani 1, 40127 Bologna, Italy}
\altaffiltext{10}{Space Telescope Science Institute, 3700 San Martin
Drive, Baltimore, MD 21218, USA} \altaffiltext{11}{Department of
Physics, Carnegie Mellon University, Pittsburgh, PA 15213, USA}
\altaffiltext{12}{Based on observations made at the European Southern
Observatory, Paranal, Chile (ESO programs 170.A-0788, 171.A-3045,
072.A-0139)}


\begin{abstract}

We present an analysis of 109 moderate-luminosity ($41.9\leq
\log~L_{0.5-8.0~{\rm keV}}\leq 43.7$) AGN in the Extended $Chandra$
Deep \hbox{Field-South} survey, which is drawn from 5,549 galaxies
from the \hbox{COMBO-17} and GEMS surveys having $0.4 \leq z \leq
1.1$.  These obscured or optically-weak AGN facilitate the study of
their host galaxies since the AGN provide an insubstantial amount of
contamination to the galaxy light.  We find that the color
distribution of AGN host galaxies is highly dependent upon (1) the
strong color-evolution of luminous ($M_V<-20.7$) galaxies, and (2) the
influence of $\sim10~ {\rm Mpc}$ scale structures.  When excluding
galaxies within the redshift range $0.63 \leq z \leq 0.76$, a regime
dominated by sources in large-scale structures at $z=0.67$ and
$z=0.73$, we observe a bimodality in the host galaxy colors.  Galaxies
hosting AGN at $z\ga0.8$ preferentially have bluer (rest-frame
$U-V<0.7$) colors than their $z\lesssim0.6$ counterparts (many of
which fall along the red sequence).  The fraction of galaxies hosting
AGN peaks in the ``green valley'' ($0.5<U-V<1.0$); this is primarily
due to enhanced AGN activity in the redshift interval $0.63\leq z \leq
0.76$.  The AGN fraction in this redshift and color interval is 12.8\%
(compared to its `field' value of $7.8\%$) and reaches a maximum of
$14.8\%$ at $U-V\sim0.8$.  We further find that blue, bulge-dominated
(S\'{e}rsic index $n>2.5$) galaxies have the highest fraction of AGN
($21\%$) in our sample.  We explore the scenario that the
evolution of AGN hosts is driven by galaxy mergers and illustrate that
an accurate assessment requires a larger area survey since only three
hosts may be undergoing a merger with timescales $\lesssim 1$ Gyr
following a starburst phase.

\end{abstract}





\section{Introduction}

There has been remarkable evidence found in the last few years that
the evolution of supermassive black holes (SMBHs) and galaxies are
inextricably linked.  For instance, ultraluminous infrared and
submillimeter galaxies \citep{al05}, with prodigious rates of star
formation, show a high fraction of AGN activity.  Current models of
galaxy evolution \citep[e.g.,][]{di05,hop05,sp05b} demonstrate that
merger-driven feedback from an AGN may quench star formation,
contributing to the order-of-magnitude decline in the cosmic SFR and
QSO emissivity \citep{bo98,fr99,me04} observed between $z\sim1$ and the
present day.  This self-regulated growth may then establish the
correlation between the mass of the central black hole and the stellar
velocity dispersion of the host bulge \citep[][]{fe00,ge00,tr02}.

A clear imprint of this scenario may be evident in the properties
(e.g., color and morphology) of the host galaxies of AGN caught at
early epochs in their evolution.  Much effort has been undertaken to
detect the hosts of nearby bright QSOs
\citep[e.g.,][]{ba97,ja04a,mc04} though this endeavor remains
challenging.  Surveys of obscured AGN, selected through various means
(e.g., optical emission lines, X-rays, or reradiated infrared
emission), offer the potential to measure cleanly the ages, stellar
populations, and star-formation rates of these host galaxies
\citep[e.g.,][]{st06}.  \citet{ka03} clearly show that the hosts of
narrow emission line AGN ($z<0.3$) in the Sloan Digital Sky Survey
(SDSS) are predominately massive, early-type galaxies and that those
with the most-luminous AGN have significant star formation with
younger mean stellar ages.  Deep X-ray surveys \citep[see][for a
review]{br05} are generating significant samples of obscured AGN
closer to the peak ($z\sim1$) of the global star-formation and merger
rate of galaxies \citep{hop06b,kart07}.  Recently, \citet{na07} have
shown that the host galaxies of moderate-luminosity AGN with
$0.6<z<1.4$, found in the Extended Groth Strip, have a broad range of
optical colors that span the same region of the color-magnitude
relation as luminous ($M_B<-20.5$) galaxies.  The prevalence of AGN
host galaxies within the region separating the blue and red galaxy
populations may lend support for the importance of AGN feedback since
this location is thought to represent a transitional phase in the
evolution of galaxies.  

It is of much interest to constrain observationally whether
environment plays a significant role in the growth of SMBHs as
expected by merger-driven accretion models.  Observational support for
the importance of mergers is mainly based on circumstantial evidence:
(1) AGN predominately reside in massive, early-type galaxies
\citep[e.g., ][]{ka03}, of which many are massive ellipticals, well
thought to be the end-product of a major-merger between gas-rich disk
galaxies, (2) ultraluminous infrared galaxies, that are
morphologically disturbed in almost all cases, have a high fraction
($\sim50\%$) of AGN activity and large molecular gas concentrations in
their nuclei \citep[see ][]{sa96}.  Studies of low-redshift ($z<0.3$)
AGN from the SDSS indicate that the fraction of galaxies harboring AGN
is independent of environment \citep[e.g.,][]{mi03} even in the cores
of massive clusters.  For the most-luminous narrow-line AGN in the
SDSS \citep{ka04}, an AGN-fraction--density relation has been shown to
exist analogous to the SFR-density relation with a higher AGN fraction
in low-density (i.e., `field') environments.  Both studies provide
evidence counter to the expectation that SMBH accretion is induced by
galaxy mergers.  Hard X-ray selected surveys, most effective at
identifying obscured accretion at higher redshifts, also present
disparate views on the relationship between AGN activity and their
environments.  \citet{ge07} present preliminary results that
demonstrate that X-ray selected AGN at $z\sim1$ `avoid' underdense
regions and those with blue ($U-V\lesssim1$) hosts are found in denser
environments, hinting at a connection to star formation.  On the
contrary, \cite{gr05} find that AGN in the $Chandra$ Deep Fields show
no evidence for an environmental dependency based on similar AGN host
morphologies and near-neighbor counts to the non-active galaxy
population.  The AGN fraction in clusters \citep{ma06,ma07}, though
higher than previously determined \citep{dr83}, is not significantly
different than that in the `field'; a different picture may emerge at
higher redshifts since \citet{east07} find significant evolution of
the AGN fraction in clusters at $z\sim0.6$.  There is also evidence
that larger scale ($\gtrsim$0.1~Mpc) structures may play an important
role in AGN fueling.  \citet{li06} have measured the cross-correlation
function of narrow-line AGN in the SDSS with a well-controlled sample
of non-active galaxies, and find that AGN are primarily not associated
with major mergers of galaxies, and preferentially lie within the
densest peaks of the underlying dark-matter distribution.
\citet{gi03} find that 30\% of the X-ray selected AGN in the 1 Ms
CDF-S are located within narrow redshift slices at $z=0.67$ and
$z=0.73$ and spread across the full $Chandra$ field-of-view
($17\arcmin\times17\arcmin$), which corresponds to a physical scale of
7.3 Mpc at $z=0.7$.  Similar structures are evident in the CDF-N
\citep{ba03} that may indicate the importance of large-scale ($\sim10$
Mpc) structures to trigger mass accretion onto SMBHs.

The Extended $Chandra$ Deep Field-South (\hbox{E-CDF-S}) is an ideal
survey field to use for investigating the properties of galaxies
harboring AGN and the role of environment due to its remarkable
multi-wavelength coverage.  We have completed a 1 Ms $Chandra$ Legacy
program \citep[P.I.: W. N. Brandt;][]{le05} that covers a wide area
(0.33 deg$^2$; three times area of the CDF-S) at the depths required
to detect moderate-luminosity ($L_{\rm X}\sim10^{43}$ erg s$^{-1}$)
AGN, including those with significant obscuration, out to the quasar
epoch ($z\sim2.5$).  The optical coverage of the \hbox{E-CDF-S} is
extensive, providing large galaxy samples with a wealth of
multiwavelength data.  For example, the COMBO-17 survey \citep{wo04}
has imaged the field with 12 narrow-band and 5 broad-band optical
filters thus providing magnitudes, accurate photometric redshifts
($\sigma_z \approx$ 0.03) and galaxy SED types, that further aid in
the identification of the X-ray source counterparts.  Optical
morphological information is available from the {\it HST} Advanced
Camera for Surveys (ACS) observations via the GEMS \citep{ri04,ha07},
GOODS \citep{gi04} and the {\it HST} Ultra Deep field
\citep[UDF;][]{beck06} projects.  Over 1000 spectroscopic redshifts
are available via the CDF-S \citep{sz04}, VVDS \citep{lef04}, K20
\citep{mi05}, and GOODS \citep{va05,va06} surveys.

In this paper, we investigate the location of moderate-luminosity AGN,
in the \hbox{E-CDF-S}, on the color-magnitude diagram and their
relation to the underlying galaxy population using our current catalog
of X-ray selected AGN with either spectroscopic or photometric
redshifts.  The \hbox{E-CDF-S} contains two prominent redshift spikes
\citep{gi03} enabling us to determine the influence of $\sim$~10~Mpc
structures on the overall color-magnitude distribution.  We discuss
how our results fit in with the morphological properties of the sample
and the impact on galaxy evolution models that incorporate AGN
feedback to quench star formation effectively.  Throughout this work,
we assume $H_0=70$ km s$^{-1}$ Mpc$^{-1}$, $\Omega_{\Lambda}=0.7$, and
$\Omega_{\rm{M}}=0.3$.

\section{Data}

We select a parent sample of galaxies in the \hbox{E-CDF-S} using
published catalogs from the COMBO-17 and GEMS surveys.  The $Chandra$
observations, covering the equivalent sky area, enable identification
of those galaxies that harbor moderate-luminosity AGN in a manner that
is least biased against obscuration.  Follow-up optical spectroscopy
of these X-ray sources with the VLT facilitates the identification of
AGN.  As detailed below, our selection is tuned to generate a sample
of galaxies hosting AGN for which the optical emission is dominated by
the host galaxy; thus no further removal of AGN light (i.e., cleaning)
is required for this study.  By restricting ourselves to an initial
optically-selected sample of galaxies, we aim to measure the fraction
of galaxies harboring AGN as a function of their intrinsic properties.

\subsection{Parent galaxy population}

COMBO-17 provides a highly complete sample of galaxies over the full
\hbox{E-CDF-S} \citep{wo04} with well-known intrinsic properties
(i.e., magnitudes and colors).  The survey provides reliable object
classifications (e.g., Galaxy, QSO, Star) and photometric redshifts by
fitting synthetic template optical spectra to the observed magnitudes
over a broad wavelength range (3500--9300 {\rm \AA}).  The source
catalog contains 8,565 objects with aperture magnitudes
$R_{ap}\leq24$, a limit at which there are photometric redshift errors
of $\delta_z/(1+z) < 0.1$.  This magnitude limit ensures that our
sample is representative of galaxies of all colors with $M_V\lesssim
-21$ and $z<1$, a limit that will be shown to be relevant for this
study.  Full details including source detection, photometry, object
classification, and redshift estimation can be found in \citet{wo04}.
Morphological information is provided by {\it HST}/ACS imaging of the
field in both the F606W and F850LP filters (hereafter referred to
$V_{606}$ and $z_{850}$, respectively) from GEMS.  Here, we use the
S\'{e}rsic indices ($n$) given in \citet{ha07} to discriminate between
bulge and disk-dominated galaxies.

We select a sample of 5,549 galaxies with photometric redshifts of
$0.4\leq z\leq 1.1$ from the GEMS $z_{850}$-band selected catalog of
\citet{ca05} that has been cross-referenced to objects in COMBO-17
($R_{ap}\leq24$).  This ensures that the sample can be used for
morphological studies and allows an upper limit to be placed upon the
optical contribution from each AGN to the color of its host galaxy.
The sample includes only those objects that are classified as a
`Galaxy' and excludes those identified as a `QSO'.  This selection
effectively avoids the presence of luminous type 1 AGN that severely
dilute the emission from their host galaxies.  The lower redshift
limit is set to 0.4 because this volume within the \hbox{E-CDF-S} is
too small at lower redshifts to generate statistically useful samples
of luminous ($M_V\lesssim-21$) galaxies and hence those with AGN.
Only two AGN have been unambigously identified at $z<0.4$ that have
$L_{\rm X}\sim10^{42}$ erg s$^{-1}$.  The redshift cutoff at 1.1 is
the limit for which COMBO-17 no longer provides accurate source
classification, and redshifts are susceptible to large uncertainties.
We primarily use the rest-frame optical magnitudes ($M_U$, $M_V$; Vega
magnitudes), publicly available from COMBO-17 \citep{wo04}, derived
from synthetic galaxy templates to measure intrinsic luminosity and
color.  To compare with theoretical models ($\S$~\ref{text:mergers}),
we utilize the rest-frame $u$ and $r$ (SDSS) also provided by
COMBO-17.  The errors \citep[$1\sigma$; ][]{wo04} on rest-frame
magnitudes are typically 0.11 ($M_U$) and 0.15 mag ($M_V$).
Throughout this work, a representative error ($1\sigma$) on color (U-V)
is determined to be 0.19 from the quadrature of the errors of the
individual rest-frame magnitudes; this is most likely an overestimate
since these magnitude errors are correlated.  \citet{be04} report a
typical error on U--V to be $\sim0.1$ mag.

\subsection{AGN identification}

We compile a sample of AGN, based on their high X-ray luminosities, in
the \hbox{E-CDF-S} by matching the 762 {\it Chandra} point sources
given in \citet{le05} to the available optical and near-infrared
catalogs.  We do not consider here the fainter X-ray sources solely
detected in the 1~Ms CDF-S to maintain a fairly uniform sensitivity
across the entire \hbox{E-CDF-S}.  Redshifts are available for 362
(48\%) of the X-ray sources through either spectroscopic or
photometric techniques.  We significantly improve upon the 97
spectroscopic redshifts available in the literature
\citep{sz04,lef04,mi05,va05,va06} with 95 additional redshifts
acquired by observations with VIMOS \citep{lef03} on the VLT through
the ESO programs 072.A-0139 (P.I.: J. Bergeron; 65 redshifts;
Silverman et al. 2007) and 171.A-3045 (GOODS; P.I.: C. Cesarski; 30
redshifts; Popesso et al., in preparation).  Our VIMOS observations
are unique since we have acquired spectra over a wide wavelength range
(3600--9500 ${\rm \AA}$) for many targets by observing with both the
LRblue and MRred grisms.  With integration times reaching 5 hr, we are
able to detect faint spectral features (e.g., [O II], Mg II, Fe II) in
many of the host galaxies of obscured or optically-faint AGN that
enable a redshift measurement.  A detailed discussion of the VIMOS
observations will be provided in a subsequent paper (Silverman et
al. 2007, in preparation).  The X-ray luminosity of each AGN is
determined from their observed broad-band (0.5-8.0 keV) flux, given in
\citet{le05}, with a $k$-correction based on a power-law spectrum
(photon index $\Gamma=1.9$) and no correction for intrinsic
absorption.  Here, we restrict the luminosity range to $41.9\leq
\log~L_{0.5-8.0~keV}\leq43.7$.  The lower luminosity limit is chosen
to provide a robust sample free of any galaxies having significant
X-ray binary or diffuse X-ray emission; luminous starburst galaxies in
the $Chandra$ Deep Fields are typically a factor of $\sim10$ fainter
\citep[see Fig. 4 of ][]{ba02}.  We also minimize potential
luminosity-dependent effects since lower luminosity AGN would only be
detected over a small fraction of the redshift range ($z =$~0.4--1.1)
considered here.  The $Chandra$ observations of the E-CDF-S are
capable of detecting AGN at this lower luminosity limit up to
$z\sim0.85$.  Our upper limit is motivated by \citet{si05} who show
that the optical emission, associated with X-ray selected AGN at
$\log~\nu l_{\nu}<43.3$ at $E = 2$~keV is primarily due to their host
galaxy since there is a strong departure of these AGN from the known
$l_{\rm opt}-l_{\rm X}$ relation for more luminous X-ray selected AGN.
We convert this limit to a broad-band (0.5--8.0 keV) X-ray luminosity
of $10^{43.7}$ erg s$^{-1}$ assuming the same spectrum as given above.
Furthermore, our adopted upper limit is similar to that of
\citet{na07} who demonstrate that the host galaxies of
moderate-luminosity AGN, at X-ray fluxes equivalent to those detected
in the \hbox{E-CDF-S}, contribute the majority of the total optical
emission.  As detailed below, we have measured upper limits to the AGN
contribution, using the {\it HST}/ACS imaging of the field, and show
that the optical emission, for the majority of our sample, is
dominated by the host galaxy.  Optical spectra provide a final check
on the AGN contribution with most sources clearly lacking any faint
broad emission lines or a rising blue continuum.

A cross-correlation of the X-ray and optical catalogs has indicated
that 109 of the 5,549 galaxies have AGN with $L_{\rm X} \approx
10^{41.9}$--$10^{43.7}$ erg~s$^{-1}$.  Of these, spectroscopic
redshifts were available for 54 of these sources, 16 (7 at $z>1$) of
which were from our VIMOS observations.  Rest-frame magnitudes for 12
galaxies were rederived (C. Wolf, private communication) using
COMBO-17 tools since their spectroscopic redshifts differed
substantially from their photometric redshifts ($\Delta z > 0.1$) that
rendered their rest-frame optical magnitudes published by COMBO-17 as
inaccurate.  This is not surprising since seven of them have
spectroscopic redshifts of $z\sim1$, a regime where photometric
redshifts for galaxies in COMBO-17 show a higher dispersion \citep[see
Figure 6 of][]{wo04}, and lack strong continuum features (e.g.,
4000~\AA~break) most useful for adequate photometric redshift
estimates.  Our AGN sample is primarily radio-quiet since only nine
have a radio-loudness ($R$) greater than 10 ($R \equiv L_{6~{\rm
cm}}/L_{4400~{\rm\AA}}$) based on 20cm VLA detections (Tozzi et al. in
preparation) that reach a flux limit of 8.5 $\mu Jy$, optical emission
attributed to the AGN (25\%; see the following section) based on our
$R$-band optical magnitudes, and spectral indices ($L_{\nu} \propto
\nu^{-\gamma}$; $\gamma_{radio}=0.8$, $\gamma_{opt}=0.5$).  We refer
the reader to \citet{ro07} for further details on the radio properties
of X-ray selected AGN in the E-CDF-S. We further note that an
additional 31 AGN have redshifts and luminosities within our range of
interest though fell out of the sample due to their faint optical
magnitudes ($R_{ap}>24$).  Our final sample of 109 AGN allows us to
compare the properties of their host galaxies to the underlying galaxy
population.

\subsection{AGN contribution to host-galaxy emission}
\label{contamination}

\begin{figure}

\includegraphics[scale=0.42,angle=0]{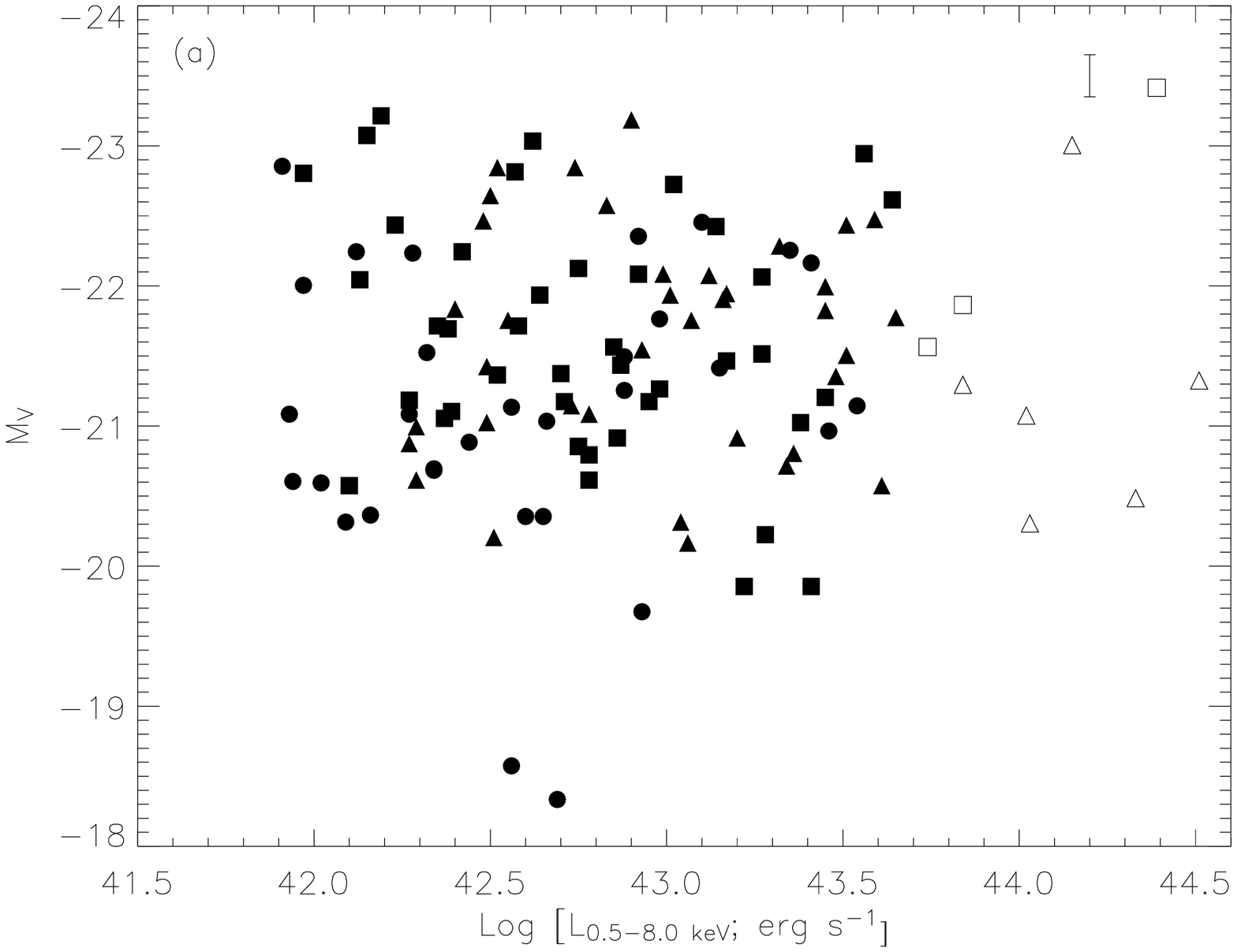}
\includegraphics[scale=0.42,angle=0]{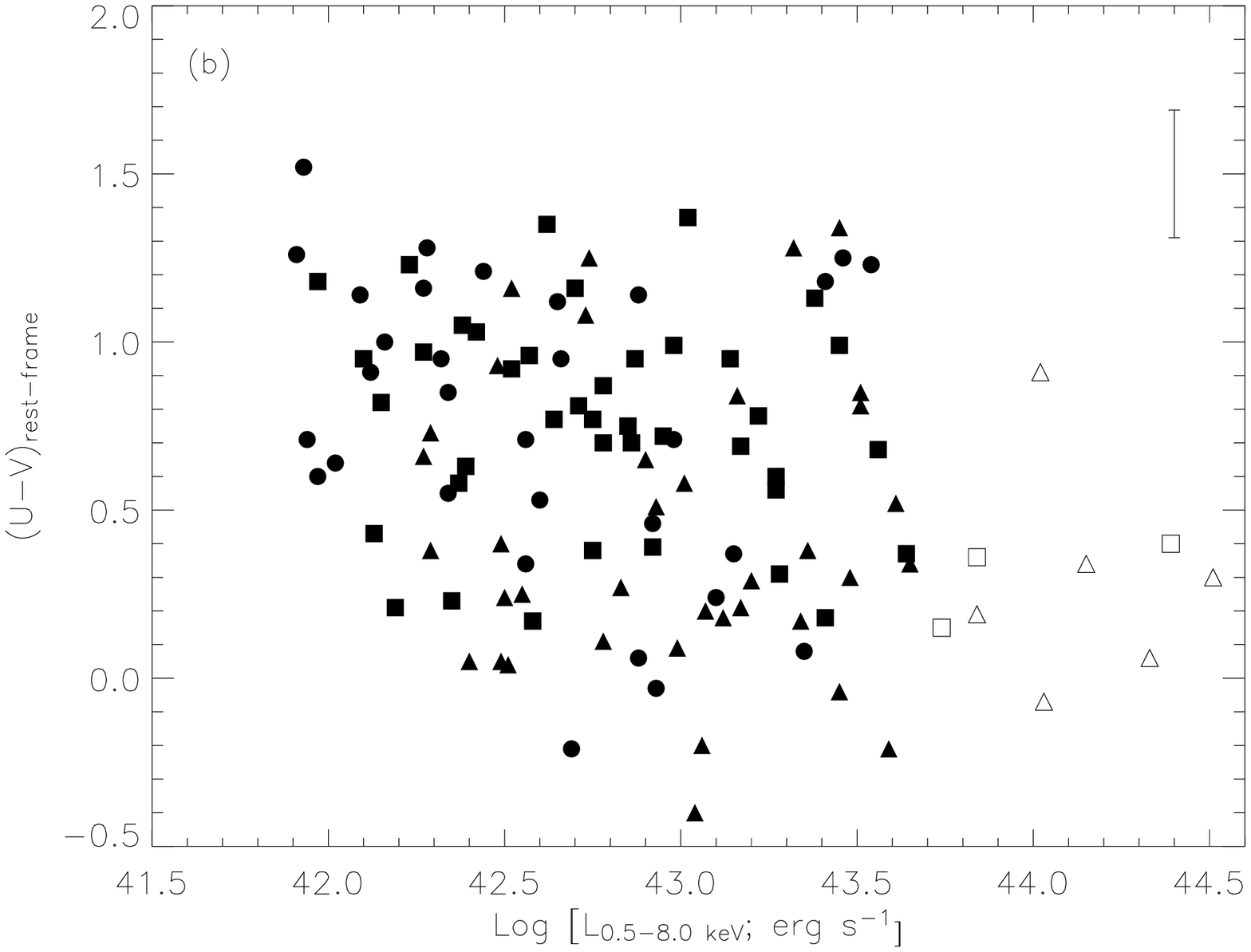}
\caption{(a) Rest-frame absolute magnitude ($M_V$) and (b) optical
color ($U-V$) as a function of X-ray luminosity for our sample of 109
AGN (filled symbols).  The type of marker denotes AGN within three
separate redshift intervals (circles: $z=$~0.4--0.63; squares:
$z=$~0.63--0.76; triangles: $z=$~0.76--1.1).  Open symbols show higher
luminosity AGN not included in our sample.  A typical error bar of
size $\pm1\sigma$ is placed in the upper right corner in both panels.}
\label{agn_sample}
\end{figure}

We are confident that the optical emission, from these 109 galaxies
hosting AGN, is dominated by star light and not strongly influenced by
the AGN based on the following arguments.  In Figure~\ref{agn_sample},
we show that there is no correlation between X-ray luminosity and
either rest-frame magnitude ($M_V$; Pearson correlation coefficient
$r=-0.036$) or color ($U-V$; $r=-0.233$).  This is consistent with the
results of \citet{na07} who show that only a weak correlation exists
for X-ray sources at similar low X-ray and optical fluxes.  The lack
of any correlation in our sample is due to the fact that we
deliberately excluded the luminous ($\log L_{\rm X} \gtrsim 44$) AGN
and any objects classified by COMBO-17 as a ``QSO''.

We determine conservative upper limits upon the AGN contribution to
the total (galaxy + AGN) light using the {\it HST}/ACS $V_{606}$ and
$z_{850}$-band images provided by GEMS.  These images, available from
MAST (Multi-mission Archive at Space Telescope), have been through the
$mark1$ processing as described in \citet{ca05}, and are rescaled to
0.03\arcsec~pixel$^{-1}$.  A few AGN are not included in our analysis
due to their proximity to the ACS field edge.  Extended optical
emission is clearly evident for all 109 galaxies hosting AGN.  Optical
counts are measured without subtracting the background in circular
apertures of two different sizes positioned at the centroid of the
optical emission: (1) a small aperture with a radius of 3 pixels
(0.09$\arcsec$) that contains $50\%$ of the flux from an unresolved
point source \citep[see][]{ja04b} and corresponds to a physical scale
of 0.48 ($z=0.4$) to 0.74 ($z=1.1$) kpc, and (2) a larger aperture
with a radius of 25 pixels (0.75$\arcsec$; covering a physical scale
of 4.03--6.13 kpc).  The size of the larger aperture is set to that
implemented for flux extraction by COMBO-17 \citep{wo04}.  The counts
within the small aperture provide a firm upper limit to the AGN
contribution since we make no attempt to remove emission of stellar
origin.  For comparison, we measure counts in equivalent regions for
the entire sample of 5,549 galaxies and 67 QSOs, which were identified
by COMBO-17 and are not included in our parent galaxy sample.

The number distribution of the AGN host galaxies as a function of the
ratio of counts between these two apertures is shown in
Figure~\ref{contamination} for both the $V_{606}$-band (Fig.~2a) and
$z_{850}$-band (Fig.~2b).  Based on the $V_{606}$-band measurements,
the mean ratio of counts of our AGN sample (solid histogram) is $0.14$
and $81\%$ of them have a ratio less than 0.2.  The distribution is
similar to that of the galaxies (dotted histogram), though shifted
slightly by 0.04 (the difference of their median values), most likely
due to the presence of optically-faint AGN.  The host galaxy
distribution is noticeably offset from that of the optically-selected
QSOs, identified by spectral template fitting with no preference for
point-like sources \citep{wo04}, that have a mean ratio of 0.48, and
60\% of them have a ratio greater than 0.5 most indicative of
unresolved point sources.  A number of low count ratio QSOs are
evident due to the fact that COMBO-17 can recognize lower luminosity
Seyfert-1 galaxies as QSOs if strong AGN features are present.  
Similar results are found for the $z_{850}$-band measurements, which
suggests that color gradients are dominated by the host galaxies and
not the underlying AGN.  Since our sample contains a large fraction of
obscured AGN, there is a slightly larger contribution from AGN
emission in the $z_{850}$ band compared to the $V_{606}$ band, whereas
the opposite is true for the quasars.

\begin{figure}
\includegraphics[scale=0.55,angle=0]{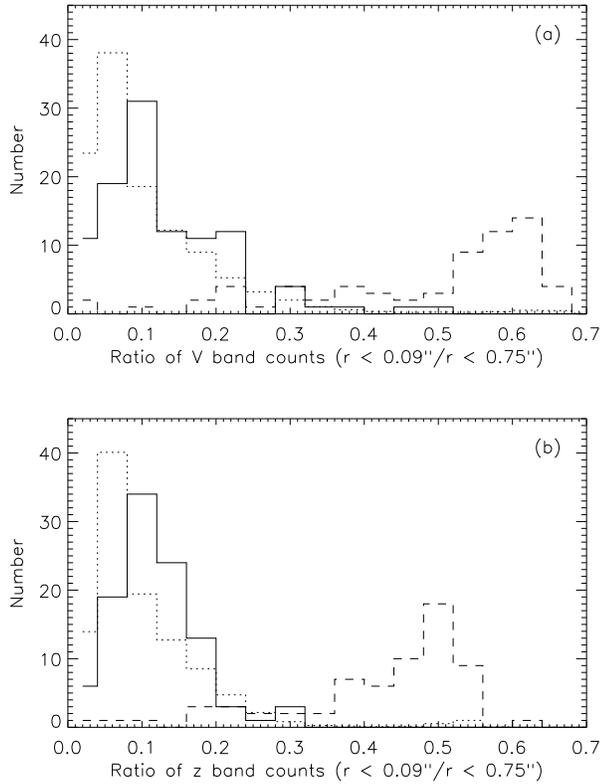}
\caption{AGN contribution to the total (AGN + galaxy) light.  The
abscissa is the ratio of source counts in {\it HST}/ACS images between
a circular aperture of radius 0.09\arcsec~and 0.75\arcsec.  In panel
a, the $V_{606}$-band count-ratio distribution is shown for 104 AGN
(solid histogram), 5521 galaxies (dotted), and 67 (dashed)
optically-selected QSOs from COMBO-17.  In panel {\it b}, the same is shown
for the $z_{850}$-band (103 AGN, 5513 galaxies, 66 QSOs).  QSOs from
COMBO-17 illustrate the typical count ratios for unresolved point
sources.  The galaxy number distribution in both panels has been
scaled down to match the numbers of AGN host galaxies.  The majority
of the AGN host galaxies have $\gtrsim 80\%$ of their total optical
emission (in both the $V_{606}$ and $z_{850}$ bands) outside the
circular aperture of radius 0.09\arcsec\ and have count ratios similar
to the non-AGN galaxy population.}
\label{contamination}
\end{figure}

The maximum amount that an AGN could shift its host-galaxy rest-frame
$U-V$ color bluewards is estimated.  We assume that an AGN contributes
all the flux in the small ($r=0.09\arcsec$) aperture, the count ratio
as detailed above is 0.2, and the background signal is negligible.
This corresponds to an AGN contributing an additional amount of flux
equal to 0.25 times the total host galaxy flux.  We further consider
an AGN to add flux to the $U$-band only.  While rest-frame UV emission
is substantially diminished at $\lambda<4000~{\rm \AA}$ for a typical
galaxy, broad-line AGN \citep{vanden06,ga06} have a strong rise in
their continuum at these wavelengths effectively driving their color
bluewards with increasing contrast between the AGN and its host-galaxy
emission as evident in Figure~\ref{contamination} with the $V_{606}$
band distribution in panel {\it a} shifted higher than that in panel
{\it b}.  We note that the rest-frame $U$-band is observed with the
$V_{606}$ band for objects with $z\sim0.7$.  Based on these
assumptions, color offsets for 81\% of our AGN sample are within
\hbox{0.24} mag.  Furthermore, we do not expect our AGN to reach this
value since a significant amount of stellar emission is present in all
cases within the small aperture and AGN light is also likely to be
emitted at longer wavelengths.  We note that if color offsets do reach
this level, the contamination is still insufficient to displace AGN
hosts from one region to another (e.g, ``green valley'' to the red
sequence).  It is more likely that the AGN contribute an additional
$\sim4\%$ to the total host galaxy flux based on the shift between the
median count ratio of the AGN host galaxies and the full galaxy
population described above.  In $\S$~\ref{discussion}, we utilize
optical spectra, which are available for 46 AGN, to further confirm
that the AGN host galaxies provide most of the optical light (see
Fig.~\ref{examples_sf} and~\ref{examples_valley}), thus supporting our
general conclusions.

\section{Rest-frame colors of moderate-luminosity AGN}

We were motivated by recent studies \citep{sa04,bo06,na07} to
investigate further the color-magnitude (rest-frame $U-V$ versus
$M_V$) relation of galaxies hosting moderate-luminosity X-ray selected
AGN.  Since the host galaxies in our sample contribute most of the
optical emission, as demonstrated in $\S$~\ref{contamination}, we do
not need to remove the AGN component from the total (host + AGN)
optical emission.  We use the rest-frame $U-V$ (i.e., $M_U-M_V$) color
up to $z\sim1$, following \citet{be04}, to quantify the color
distribution with a greater sensitivity to the continuum slope across
the 4000 ${\rm \AA}$ break than is provided by rest-frame $U-B$ color.
We note that \citet{wo04} caution that the rest-frame measurement of
$M_V$ at $z>0.7$ is based on an extrapolation of the best-fit galaxy
template outside of the observed optical passbands in COMBO-17.  For
all results presented here, we have confirmed that the same basic
conclusions are drawn by using the $M_U$ and $M_B$ rest-frame absolute
magnitudes, thus removing the possibility that our results are
dependent on the assumed spectral template (i.e., $k$-correction).

\begin{figure*}

\includegraphics[scale=0.70,angle=90]{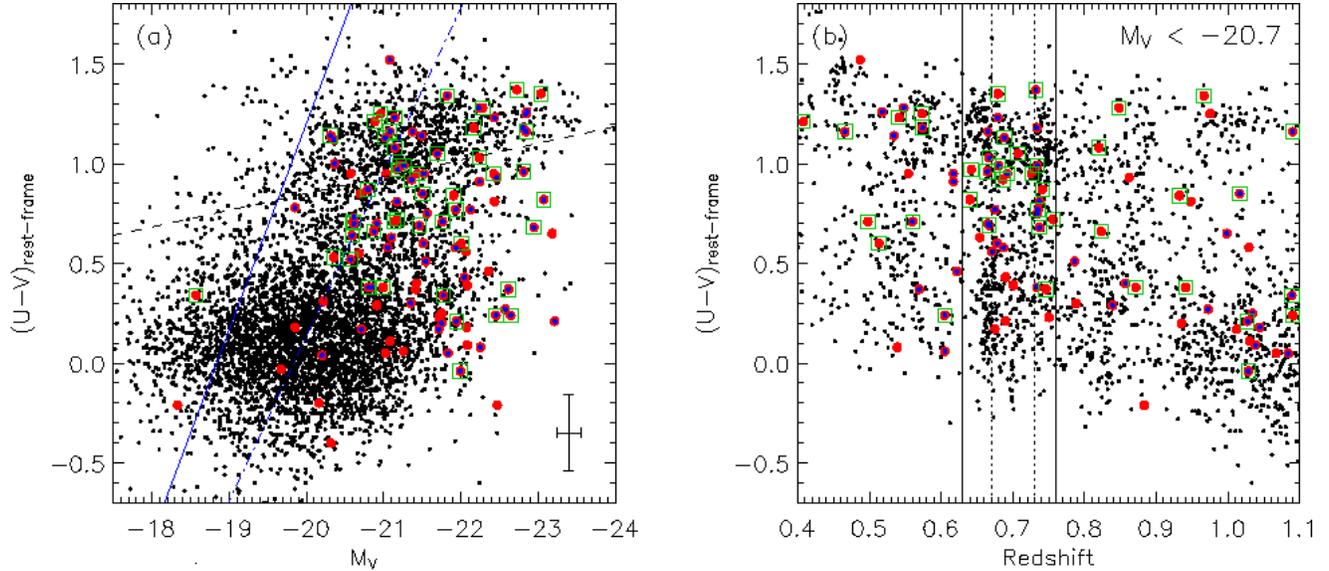}
\caption{(a) Rest-frame optical color ($U-V$) as a function of
rest-frame absolute magnitude ($M_V$) for 109 galaxies hosting AGN
compared to their parent sample of 5,549 galaxies with $0.4\leq z \leq
1.1$.  Galaxies in the parent sample are marked with small black
filled circles; those hosting X-ray selected AGN are highlighted by
large red circles.  The slanted blue lines denote the approximate
limits for galaxies with $R_{ap}\le24$ at $z=0.8$ (solid) and $z=1$
(dash-dotted).  The division between red and blue galaxies implemented
by \citet{be04} is shown by the dashed line.  AGN hosts classified as
bulge-dominated (S\'{e}rsic index 2.5--8) galaxies by the {\it
HST}/ACS morphology \citep{ha07} are marked with blue filled circles
at the centers of the red circles.  AGN with X-ray hardness ratios
larger than 0.2 (see $\S$~\ref{text:mergers}) are marked by open green
squares in both panels.  The typical error ($\pm1\sigma$) bar is shown
in the bottom right corner.  (b) Rest-frame $U-V$ color versus
redshift for the 2,044 most luminous ($M_V<-20.7$) galaxies.  The
vertical solid lines denote the redshift interval $0.63\leq z \leq
0.76$ with the dotted lines marking the redshift spikes at $z=0.67$
and $z=0.73$.  In this panel only, smaller blue dots mark the AGN with
spectroscopic redshifts.}

\label{color_mag}
\end{figure*}

In Figure~\ref{color_mag}a, we plot the rest-frame colors ($U-V$) as a
function of absolute $V$-band magnitude ($M_V$) for our galaxies,
including those hosting AGN in the redshift interval $0.4\leq z \leq
1.1$.  We confirm past results \citep{ba03,bo06,na07} with better
statistics: (1) a high fraction (80\%) of moderate-luminosity AGN
reside in the most-luminous ($M_V<-20.7$) galaxies, (2) the rest-frame
colors of AGN host galaxies have a broad distribution, over the range
$0 < U-V < 1.5$, with no apparent evidence for a color bimodality, as
is distinctively evident in the underlying population of galaxies
\citep{be04}, and (3) the majority (60\%) of AGN host galaxies have
bulge-dominated morphologies \citep[S\'{e}rsic index $n>2.5$;
][]{bl03,mc05} as marked by small blue dots.  Here, we further find
that 31--44\% of the AGN with $M_V<-20.7$ have blue colors, depending
on the chosen division ($0.6<U-V<0.8$) between blue and red galaxies
\citep[dashed line in Fig.~\ref{color_mag}a; see][]{sa04}, thus
associating them with star-forming galaxies.  This is not surprising
since we have a fair number of AGN at $z>0.8$, an epoch for which the
mean star-formation rate of galaxies has increased by an order of
magnitude compared to the present value
\citep[e.g.,][]{ho06,no07,zh07}.  These results are not strongly color
biased since both red and blue luminous ($M_V<-20.7$) galaxies, out to
$z\sim1$, mainly fall above the magnitude limits
\citep[$R_{ap}\lesssim24$; ][]{be04} shown by the blue lines in
Figure~\ref{color_mag}a ($z=0.8$: solid; $z=1.0$: dash-dotted).  We
note that some red ($U-V>1.0$), less-luminous ($-20.7<M_V<-21.3$)
galaxies at $z>0.9$ may be missed.

\begin{figure*}
\includegraphics[scale=0.65,angle=90]{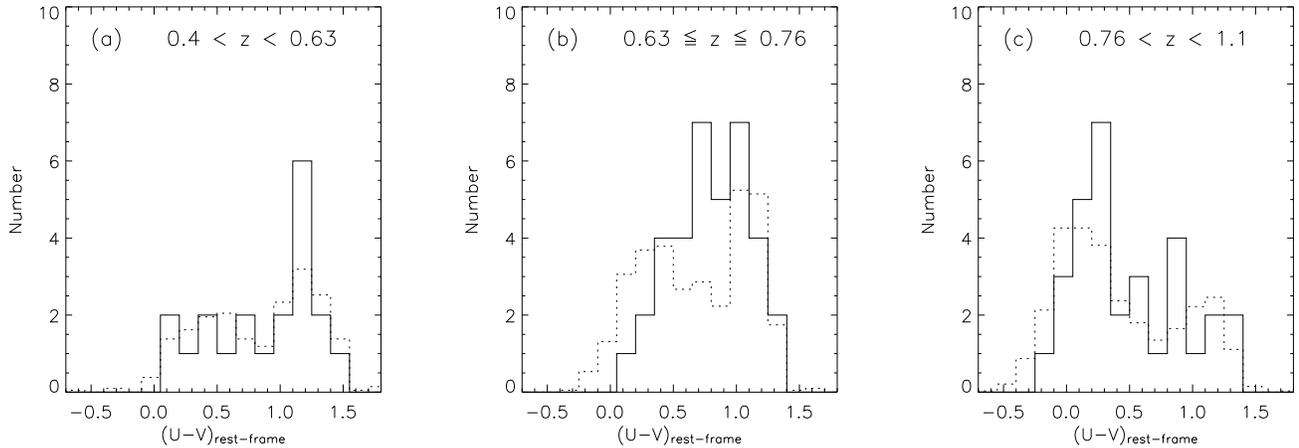}
\caption{Rest-frame color ($U-V$) histogram of galaxies that host AGN
(solid line) with absolute magnitude $M_V<-20.7$ in three redshift
intervals: $0.4<z<0.63$ (a), $0.63\leq z \leq0.76$ (b), $0.76<z<1.1$
(c).  For comparison, we have plotted, in all panels, the distribution
of galaxies (dotted line) above this absolute magnitude and
renormalized to match the number of AGN.  Note that the peak of the
distribution shifts from red to blue with increasing redshift for both
galaxies hosting AGN and the underlying galaxy population.}
\label{color_histo}
\end{figure*}

It is apparent that the color distribution of these AGN clearly
depends on redshift as shown in Figure~\ref{color_mag}b.  Here, we
only consider luminous galaxies ($M_V<-20.7$) to minimize any color
bias.  In the redshift interval $0.4<z<0.63$ (see
Fig.~\ref{color_histo}a), the AGN tend to have red colors ($U-V>0.7$)
with a mean color $\langle U-V \rangle$ = $0.86\pm0.10$.  Within the
same redshift interval, the galaxies without AGN are also
preferentially red.  We find that 70\% (14 of 20) of the AGN and 62\%
(239 of 386) of the galaxies have these colors with most associated
with the red sequence.  In contrast, the mean color of AGN hosts in a
higher redshift interval ($0.76<z<1.1$; Fig.~\ref{color_histo}c) is
$\langle U-V \rangle$ = $0.49\pm0.08$ with 29\% (9 of 31) of the AGN
having $U-V>0.7$.  A similar fraction (27\%; 211 of 790) of luminous
galaxies at these redshifts has these colors.  The variance of each
color distribution is similar ($s^2\sim0.2$).  Again, it is as
expected that the majority of AGN host galaxies at $z>0.76$ have blue
colors ($U-V<0.7$) since blue galaxies dominate the luminous
population at these redshifts \citep{wo03,be04}.  The null hypothesis,
from a Kolmogorov-Smirnov (K-S) test \citep[see ][]{pr93}, has a
probability of 2.4\% that the distribution of host galaxy colors in
Figure~\ref{color_histo}a and~\ref{color_histo}c could be drawn from
the same parent population.  We have implemented further K-S tests to
determine whether the AGN distributions could be drawn from the
underlying galaxy population.  In Table~\ref{stats}, we give the
results that show that the color distribution of AGN host galaxies in
Figure~\ref{color_histo}a ($P_{\rm K-S}=0.79$) and \ref{color_histo}c
($P_{\rm K-S}=0.21$) resembles that of the overall galaxy population.
{\it We conclude, based on these tests, that the host galaxies of
moderate-luminosity AGN follow a similar passive evolution, or aging,
as the underlying galaxy population migrates from blue to red colors
with cosmic time}.  We note that AGN with spectroscopic redshifts,
shown by a small blue dot in Figure~\ref{color_mag}b, confirm this
trend.

\begin{deluxetable*}{lllrc}
\tabletypesize{\small}
\tablecaption{Statistical comparison of color distributions\label{stats}}
\tablewidth{0pt}
\tablehead{\colhead{Note}&\colhead{\# AGN}&\colhead{\# galaxies}&\colhead{K-S}&\colhead{Redshift range}}
\startdata
AGN-galaxy (Fig.~\ref{color_histo}a)&20&415&79.4\%&$0.4<z<0.63$\\
AGN-galaxy (Fig.~\ref{color_histo}b)&36&669&6.6\%&$0.63 \leq z \leq 0.76$\\
AGN-galaxy  (Fig.~\ref{color_histo}c)&31&957&20.6\%&$0.76 < z < 1.1$\\
AGN (Fig.~\ref{color_histo}a)-AGN (Fig.~\ref{color_histo}c)&--&--&2.4\%&-----\\
AGN (Fig.~\ref{color_histo}a)-AGN (Fig.~\ref{color_histo}b)&--&--&13.4\%&-----\\
AGN (Fig.~\ref{color_histo}b)-AGN (Fig.~\ref{color_histo}c)&--&--&0.2\%&-----\\

\enddata
\end{deluxetable*}

\section{Large-scale influences}  

As previously mentioned, recent studies \citep{di05,cr06,hop06a} have
attributed the truncation of star formation and eventual redward
migration of galaxies to merger-induced AGN feedback that effectively
populates the red sequence with massive galaxies \citep{be04,fa06}.
This scenario may explain the fair number of AGN host galaxies in our
sample residing in the ``green valley'' (see Fig.~\ref{color_mag}a) if
they are preferentially located in overdense regions.

In contrast to studies that characterize the environment in terms of
density local to AGN, we are utilizing the fortuitous structures in
the \hbox{E-CDF-S} to search for large-scale effects.  To do so, we
specifically isolate a redshift interval ($0.63\leq z \leq 0.76$)
dominated by two redshift spikes \citep[Fig.~\ref{color_mag}b,
vertical dotted lines at $z=0.67$ and $z=0.73$, each with $\delta
z<0.02$;][]{gi03} evident in the 1~Ms CDF-S area, which appear to
extend over the larger \hbox{E-CDF-S} area (Silverman et al. 2007).  A
larger fraction of AGN activity in galaxies within these large-scale
structures was reported by \citet{gi03} albeit with limited
significance ($2\sigma$).  Ideally, one would like to consider even
narrower redshift intervals ($\delta z \sim 0.02$) to select galaxies
cleanly within these spectroscopic redshift spikes, but we are
restricted here by photometric redshift errors.  Approximately 80\% of
the galaxies within this redshift interval have errors in redshift
($\sigma_z$) less than 0.07\footnote{Photometric redshift errors are
based on equation 5 of \citet{wo04} and a magnitude limit of 23.5.}, a
factor of two smaller than the chosen bin width ($\Delta z=0.14$).

It is evident from Figure~\ref{color_mag}b that the majority of AGN
within this redshift interval have colors ($0.5<U-V<1.0$) placing them
in the ``green valley''.  Many of the AGN have spectroscopic
redshifts, shown by the small blue dots, that confirm their presence
within the narrow redshift spikes ($\Delta z < 0.02$) and their
rest-frame $U-V$ colors that place them in the ``green valley'' (67\%;
12 of 18 AGN).  We plot in Figure~\ref{color_histo}b the $U-V$ color
distribution of both AGN and galaxies with $M_V<-20.7$ that fall
within this redshift interval.  We have scaled the galaxy distribution
to match the number of AGN.  The color distribution of the host
galaxies is clearly different from the underlying galaxy population.
{\it We find that the host galaxies of moderate-luminosity AGN are
preferentially located at intermediate colors $(U-V\sim0.8)$ between
the red and blue galaxy populations, and skewed toward blue colors}.
A K-S test
\footnote{The results of additional tests between the color
distributions of AGN hosts (AGN-AGN) in the three redshift intervals
are reported in Table~\ref{stats}.  We note that their significance is
limited given the lack of a large comparison sample that effectively
improves the AGN-galaxy comparisons.} gives a probability of 6.6\%
that the AGN distribution could be randomly drawn from the underlying
galaxy population and suggests that the distribution of these 36 AGN
is different from that of the 669 underlying galaxies.

There does not appear to be any strong selection effects that could be
responsible for the color distribution of AGN host galaxies within
this redshift interval.  Essentially, all of the AGN have optical
luminosities well above the limit at z=0.8 shown in
Figure~\ref{color_mag}a by the solid, blue line.  By isolating
luminous ($M_V<-20.7$) host galaxies, there is clearly no color bias.
This is an important point since the $4000~{\rm \AA}$ break is moving
through the $R$-band filter at $z\sim0.7$ and red galaxies can
potentially fall below the flux limit more rapidy than those with
bluer colors.  There should not be any problems associated with the
use of COMBO-17 filters at these redshifts since both rest-frame $M_U$
and and $M_V$ fall within the observable window, and multiple filters
sample the galaxies' spectral energy distribution below and above the
$4000~{\rm \AA}$ break.  Both surface brightness dimming and the use
of a fixed photometric aperture can induce systematic changes in color
with redshift.  An increase in the contrast between a bulge and a
disk, due to surface brightness dimming as a function of redshift,
would typically redden its rest-frame color; this effect is of most
concern when comparing colors over a wider redshift baseline.
\citet{be04} address aperture effects associated with COMBO-17
photometry and state that an induced color gradient is only evident at
$z\lesssim0.4$, below the redshift range of our sample, with a
potential color offset of $\sim0.1$ mag.  There does exists the
possibility that the color distribution of AGN hosts is the result of
red and blue galaxies having roughly equivalent numbers at $z\sim0.7$,
coupled with our small AGN sample and sensitivity to cumulative biases
though our statistical test suggests otherwise.

\section{AGN fraction}

\label{text:agn_fraction}

We measure the fraction of galaxies that host AGN as a function of
color and large-scale environment.  We follow the technique discussed
in $\S$~3.1 of \citet{le07} to determine the AGN fraction for our
parent population of 5,549 galaxies.  This method properly accounts
for the spatially varying sensitivity limits of the {\it Chandra}
observations of the \hbox{E-CDF-S} \citep[see Fig.~17 of][]{le05}.
The necessity of this approach is demonstrated in
Figure~\ref{sample_limits}, which shows the limiting X-ray luminosity
as a function of redshift for the entire galaxy sample and the
measured X-ray luminosities of those galaxies harboring AGN.  In
addition to the Malmquist bias, there is almost an order-of-magnitude
spread in limiting luminosity over the full redshift range.  To
illustrate these effects, the fraction of galaxies that could host a
detectable AGN with $\log~L_{\rm X}=41.9$ is $\approx 23\%$; this
fraction rises steeply to $\approx 80\%$ when considering a limiting
luminosity of 42.5.  To account for this selection effect, we
determine the contribution of each AGN separately to the total
fraction.  The AGN fraction ($f$; see equation{~\ref{fraction} below)
and associated error ($\sigma$; see equation~\ref{fraction_error}) are
a sum over the full sample of AGN ($N$) with $N_{\rm gal,i}$
representing the number of galaxies capable of hosting the $i$th
detectable AGN with X-ray luminosity $L^{i}_{\rm X}$.

\begin{equation}
f=\sum_{i=1}^{N}\frac{1}{N_{\rm gal,i}}
\label{fraction}
\end{equation}

\begin{equation}
\sigma^2\approx\sum_{i=1}^{N}\frac{1}{{N_{\rm gal,i}^{2}}}
\label{fraction_error}
\end{equation}

\begin{figure}
\includegraphics[scale=0.42,angle=0]{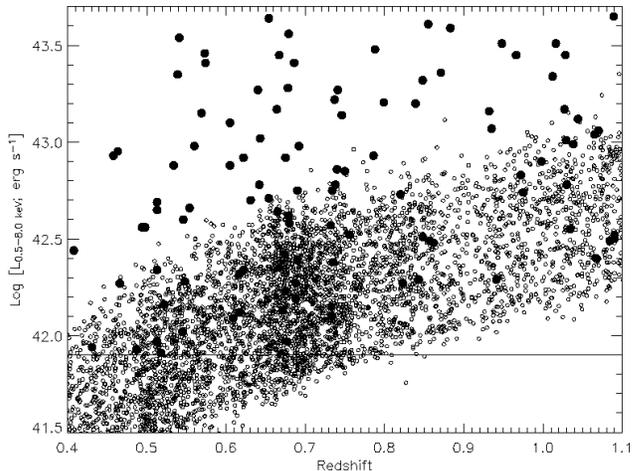}
\caption{The full band X-ray luminosity as a function of redshift for
the 5,549 galaxies in our parent sample.  The filled circles are the
X-ray detections (AGN); the open circles are upper limits.  The spread
of a factor of six in the value of the limits at a given redshift is
produced by the variations in survey sensitivity with location; for a
given redshift, the objects with the lowest limits are located at the
center of each ACIS pointing and those with the highest limits are
located at the edge of the ACIS array.  The horizontal line denotes
the minimum luminosity for inclusion in our AGN sample.}
\label{sample_limits}
\end{figure}

We have calculated the fraction of galaxies harboring AGN in bins of
color for the entire sample (Fig.~\ref{agn_fraction}a).  Using a bin
width of $\Delta (U-V) = 0.4$, we find that the fraction rises from
$\sim0\%$ at the bluest colors to $\sim5\%$ at $U-V\sim0.4$ (i.e., red
end of the blue-galaxy population), has a peak in the ``green valley''
($U-V\sim0.8$) with a value of $\sim10\%$, and then declines to
$\sim4-5\%$ at $U-V\sim1.3$ (i.e., along the red sequence).  In
Table~\ref{table_fraction}, we list the fraction of luminous
($M_V<-20.7$) galaxies hosting AGN that we measure for various
subsamples of galaxies.  We note that the AGN fraction in the `field'
(i.e., $0.4\leq z<0.63$ and $0.76<z \leq 1.1$) is $5.9 \pm 1.0\%$ (51
AGN from effectively 1157 galaxies) when considering all colors in one
bin.  Remarkably, the distinct enhancement of the AGN fraction in the
``green valley'' is due to an overdensity of AGN in the redshift
interval (0.63--0.76) that contains two prominent redshift spikes.  In
Figure~\ref{agn_fraction}b, we show the fraction of galaxies hosting
AGN in (solid line) and out (dotted line) of this narrow redshift
interval.  For an intermediate color interval $0.5<U-V<1.0$, the AGN
fraction within this narrow redshift range (solid line) is
$12.8\pm2.9\%$ (21 AGN, 173 galaxies), that reaches $\sim15$\% at
$U-V\sim0.8$, while those in the `field' (dotted line) have a fraction
of $7.8\pm2.5\%$ (14 AGN, 246 galaxies).  The significance of this
higher AGN fraction in large-scale stuctures is reported here at the
$2.6\sigma$ ($>99\%$) confidence level based on the difference and
errors of these measurements.  This analysis improves upon the
significance of enhanced AGN activity within large-scale structures
previously reported by \citet{gi03}.  A less conspicuous peak is
apparent in the `field' sample though not sufficiently significant
given the current sample size.  We find an overall higher AGN fraction
in the 'field', compared to \citet{na07}, by a factor of $\sim$1.8
though a similar $relative$ fraction as a function of color
(Fig.~\ref{agn_fraction}b; dotted line); the AGN fraction appears to
be rather flat ($\sim6\%$) across a wide range of color ($0.3 \lesssim
U-V \lesssim 1.5$) that includes the red sequence and the valley but
then drops off for blue galaxies beyond the top ($U-V<0.2$) of the
`blue cloud'.

\begin{figure}
\includegraphics[scale=0.67,angle=90]{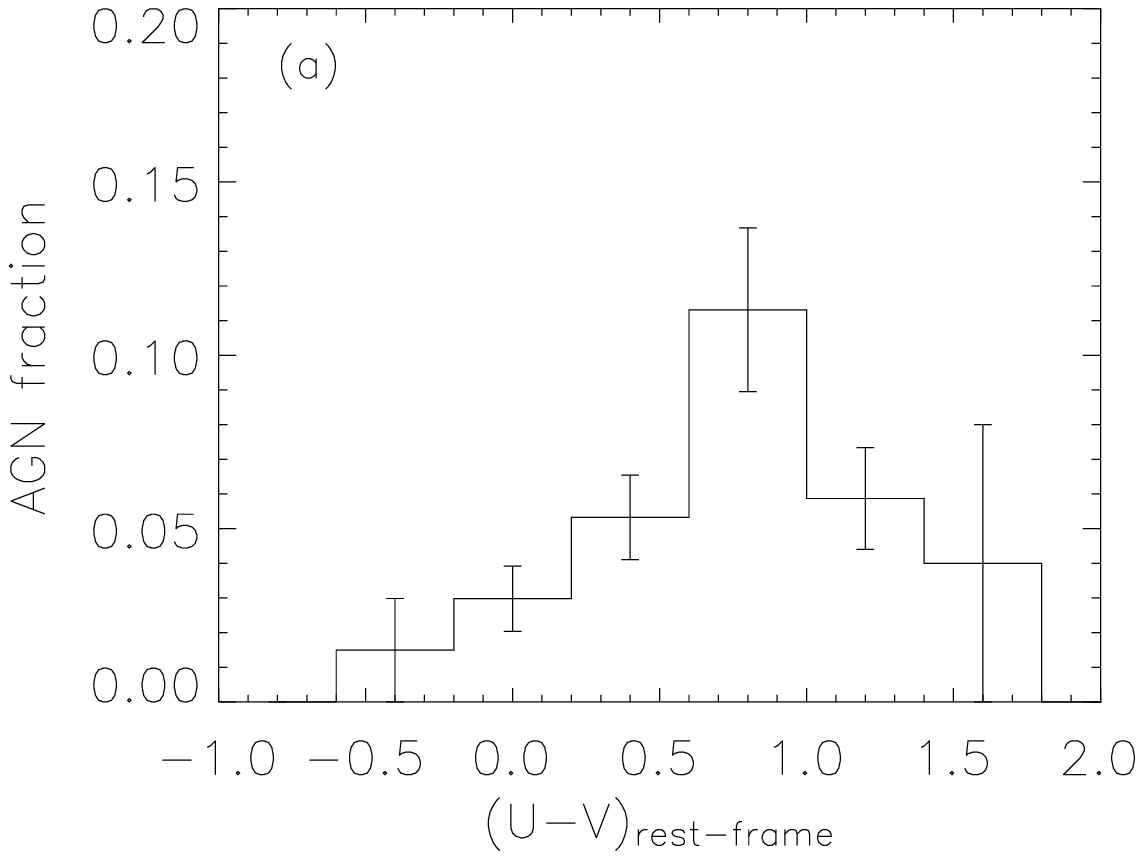}
\includegraphics[scale=0.67,angle=90]{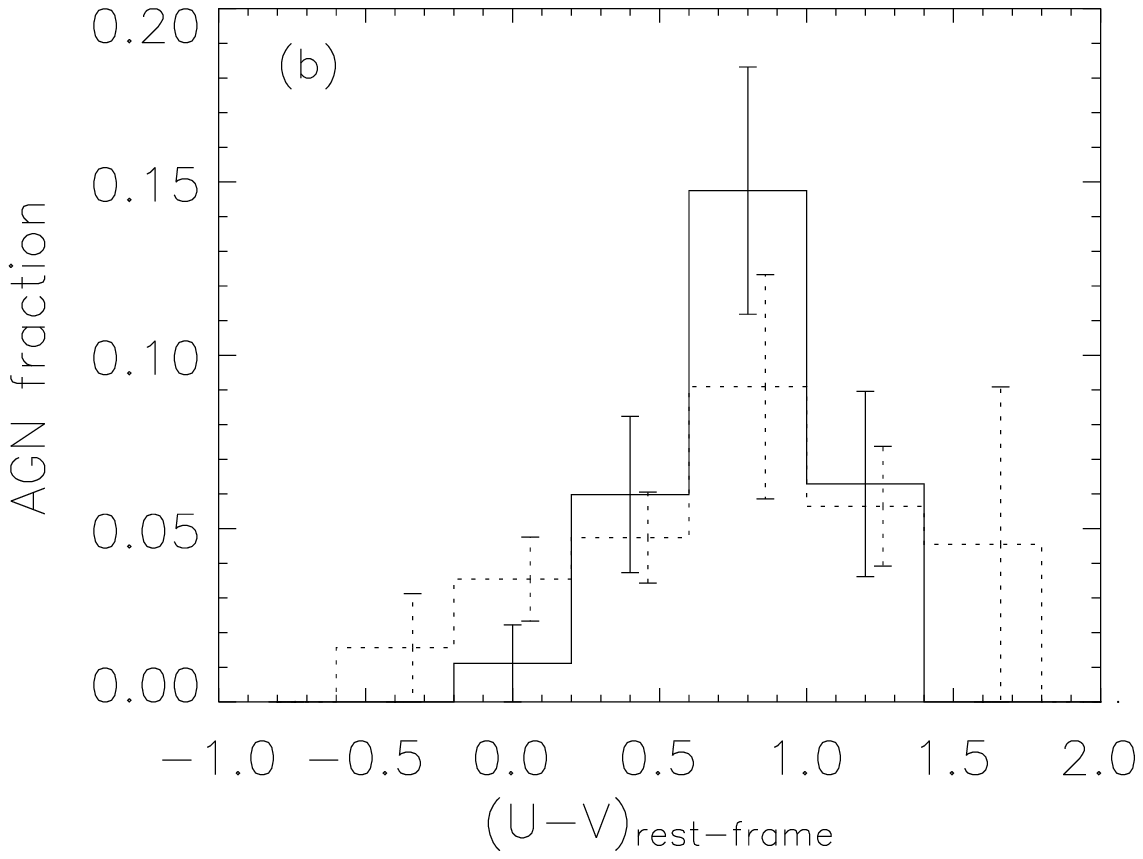}
\caption{Fraction of galaxies ($M_V<-20.7$) hosting AGN as a function
of their rest-frame optical color ($U-V$) for the entire sample (a).
In panel $b$, we have split the sample into those in (solid) and out
(dashed) of the redshift interval $0.63 \leq z \leq 0.76$.  Errors
(1$\sigma$) are shown for all values with a slight displacement of the
dashed set for visual purposes only.  A suggestive enhancement
($\sim$~two~$\times$ with significance at the $2.6\sigma$ level) of AGN
activity is evident for host galaxies residing in the ``green valley''
that is mainly attributed to activity in large-scale structures.}
\label{agn_fraction}
\end{figure}

\begin{deluxetable*}{rllccl}
\tabletypesize{\small}
\tablecaption{Fraction of luminous ($M_V<-20.7$) galaxies hosting AGN\label{table_fraction}}
\tablewidth{0pt}
\tablehead{\colhead{Index}&\colhead{Redshift\tablenotemark{a}}&\colhead{$U-V$}&\colhead{S\'{e}rsic}&\colhead{Fraction}&\colhead{Notes}\\
&\colhead{range}&\colhead{range}&\colhead{indices}&\colhead{\%}}
\startdata
1&I+II+III&---&---&$6.4\pm0.8$&Full sample\\
2&I+II+III&$<0.5$&---&$4.6\pm0.9$&Blue cloud\\
3&I+II+III&0.5--1.0&---&$9.8\pm1.9$&``Green valley''\\
4&I+II+III&$>1.0$&---&$6.0\pm1.5$&Red sequence\\
5&I+III&---&---&$5.9\pm1.0$&`Field sample'\\
6&I+III&0.5--1.0&---&$7.8\pm2.5$\\
7&II&0.5--1.0&---&$12.8\pm2.9$&Redshift spikes\\
8&I+II+III&---&0--2.5&$1.9\pm0.5$&disks\\
9&I+II+III&---&0--1.5&$1.0\pm0.4$&purely disks\\
10&I+II+III&---&2.5--8&$10.6\pm1.7$&bulges\\
11&I+II+III&$<0.7$&2.5--8&$21.3\pm5.0$&blue spheroids\\
12&I+II+III&$>0.7$&2.5--8&$7.9\pm1.7$&red spheroids\\

\enddata
\tablenotetext{a}{I: $z=$~0.4--0.63; II: $z=$~0.63--0.76; III: $z=$~0.76--1.1}
\end{deluxetable*}

\section{Discussion}
\label{discussion}
\subsection{Global evolution of host-galaxy colors}

It is useful to ask what would be the observed color distribution of
the hosts of AGN as a function of redshift up to $z\sim1$ if we simply
assume that AGN reside in the most luminous galaxies, and hence the
most massive, as corroborated by many findings
\citep[e.g.,][]{ka03,be05}.  Based on the luminosity function of
galaxies from the \hbox{COMBO-17} \citep{wo04} and DEEP2 \citep{fa06}
surveys, it is evident \citep[Fig.~6 of][]{fa06} that luminous
($M_B< -21$) galaxies at $z\lesssim0.6$ are predominately red (i.e.,
early-type).  At $z\gtrsim0.6$, a transition occurs where the majority
of luminous galaxies are blue (i.e., late-type, star-forming) as a
result of (1) the order-of-magnitude increase in the mean
star-formation rate \citep[e.g., ][]{ho06}, across all mass scales
\citep[e.g.,][]{no07,zh07}, and (2) strong depletion of galaxies along
the red sequence \citep{be04,be07} up to $z\sim1$. In
Figure~\ref{galaxy_distributions}, we show the distribution of
rest-frame $U-V$ colors for redshift-selected galaxy populations in
our sample.  The blue peak ($U-V\sim0$) of the luminous ($M_V<-20.7$)
galaxy distribution is most prominent in the highest redshift bin
($z=$~0.76--1.1; dotted histogram), while the red peak is more fully
populated in the lowest redshift bin ($z=$~0.4--0.63; dashed
histogram).

We plausibly expect to see a similar redshift dependence in the color
distribution of galaxies hosting AGN with redshift.  Observational
evidence does exist to support this scenario.  Host-galaxy studies,
pioneered by {\it HST}, show that early-type galaxies represent the
majority of the hosts of QSOs at low redshifts \citep[$z<0.3$;
e.g.,][]{ba97}.  At similar redshifts, \citet{ka03} demonstrate that
the hosts of over 20,000 narrow line AGN from the SDSS are mainly
early-type, bulge-dominated systems.  The majority of the most
luminous, type II QSOs from the SDSS \citep{za06} and radio galaxies
\citep{mc04} appear to also have similar early-type hosts.  At higher
redshifts ($z>0.5$), there is now evidence that a large fraction of
AGN hosts have blue colors and hence significant star formation
\citep{ja04b,sa04,wi06}.  However, there is little evidence that the
host galaxies of AGN resemble the blue, star-forming galaxies that
dominate the galaxy population at $z>0.8$ as pointed out by
\citet{ba03} in reference to X-ray selected samples.

\begin{figure}
\includegraphics[scale=0.67,angle=90]{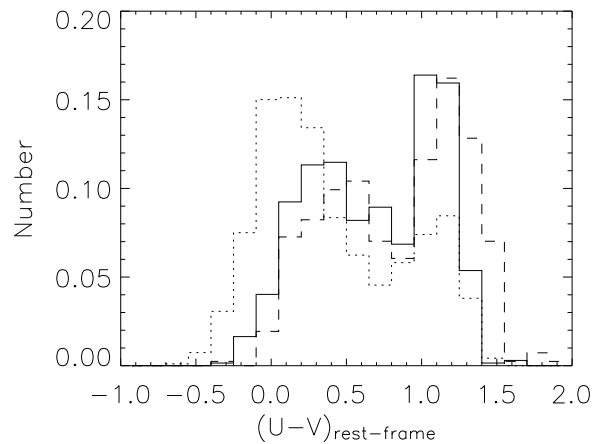}
\caption{Color distribution of all luminous ($M_V<-20.7$) galaxies in three
redshift intervals (dashed: $z =$~0.4--0.63; solid: $z =$~0.63--0.76; dotted:
$z =$~0.76--1.1).  All have been normalized by the number of objects in each
interval.}
\label{galaxy_distributions}
\end{figure}

In our sample, the color distribution of luminous ($M_V<-20.7$) AGN
host galaxies in the `field' (i.e., outside the redshift interval
containing prominent redshift spikes) appears to support the above
scenario.  In Figure~\ref{color_histo}, we see that the distribution
(solid histogram) has both a red (Fig.~\ref{color_histo}a) and a blue
(Fig.~\ref{color_histo}c) peak for the $0.4 < z < 0.63$ and $0.76 < z
< 1.1$ redshift bins, respectively; these peaks are coincident with
those of the underlying galaxy populations (dotted histogram).  Since
the AGN sample is limited in size, this bimodality is only evident
when considering AGN over a wide redshift range and not at each epoch
as characteristic of the color distribution of large samples of
galaxies \citep[e.g.,][]{st01,bl03,ba04,ba06,ca07}.  The colors of the
host galaxies at $z\lesssim0.6$ are predominately red ($U-V>0.7$;
Figure~\ref{color_mag}b) and lie on or close to the red sequence.
Optical spectra are available for five of the reddest ($U-V>1.0$) AGN
host galaxies and each has a strong 4000~\AA~break characteristic of
an old stellar population.  At higher redshifts ($z\gtrsim0.8$), the
AGN\footnote{A handful appear to be associated with a less prominent
redshift spike identified by \citet{gi03} at $z\sim1.04$.} are more
prevalent in blue (i.e., star-forming) galaxies.  Nine of these
high-redshift AGN with $U-V<0.4$ have optical spectra that confirm
that their colors are dominated by stellar emission.  In
Figure~\ref{examples_sf}, we show representative optical spectra from
our VLT program that are characterized by weak 4000~\AA~breaks, fairly
flat blue continua, and no optical signatures of an embedded AGN in
all but one case (Source \# 333).  To summarize, we find that the
color distribution of AGN host galaxies has a redshift dependency that
reflects that of the underlying luminous galaxy population.  It is
also apparent that AGN in the `field' (Fig.~\ref{agn_fraction}b;
dotted line) are equally likely to reside in galaxies over a broad
range in color with the exception of those at the blue ($U-V<0.2$)
end.  This result suggests that AGN activity in the `field' is
primarily dependent on mass\footnote{Using equation 2 of
\citet{bell05}, we confirm that estimates of stellar mass of AGN host
galaxies in our sample, based on their rest-frame colors, represent
the high end of the galaxy mass distribution since we find a mean of
$\sim1.3\times10^{10}$ M$_{\sun}$ and 90\% of our sample with $M
\gtrsim 3.9\times10^{9}$ M$_{\sun}$, in agreement with
optically-selected AGN host galaxies \citep{sa04}.}, given the strong
dependency on observed luminosity and weaker in the luminous, blue
(i.e., star-forming) galaxy population.  In $\S~\ref{morphology}$, we
demonstrate that the lack of an AGN-star formation connection only
applies to disk-dominated galaxies.

\begin{figure*}
\includegraphics[scale=0.75,angle=0]{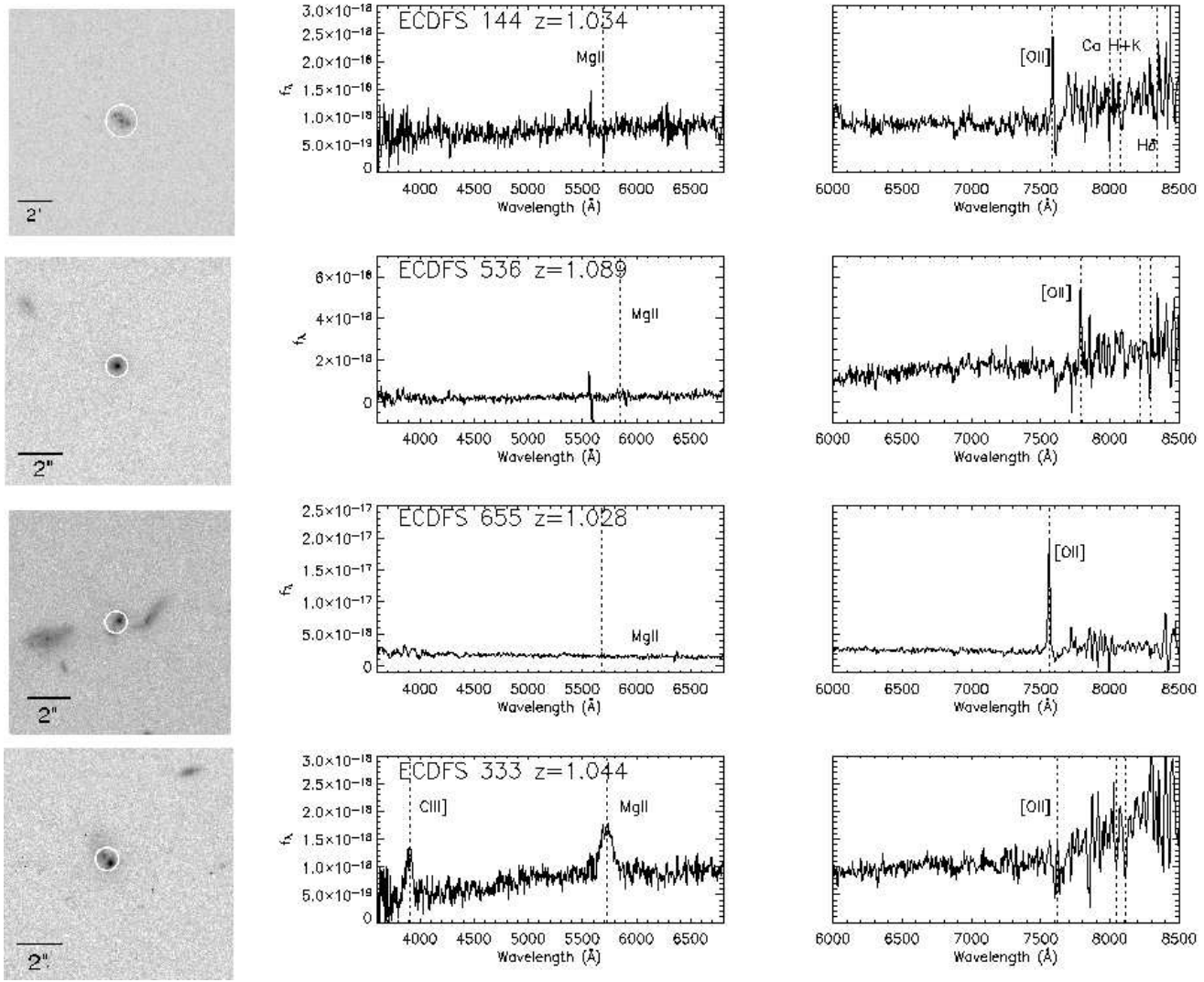}
\caption{Examples of blue ($U-V<0.4$) galaxies hosting AGN at $z>0.8$
that have been identified by our VLT/VIMOS observations.  The {\it
HST}/ACS $z_{850}$-band images are shown with a log scaling.  The optical
spectra that cover a wavelength range of 6000--8500~\AA~(right column)
have been binned by 2 pixels.  The flux scale is erg cm$^{-2}$
s$^{-1}$ ${\rm \AA^{-1}}$.  The source numbers follow those presented in
the \citet{le05} X-ray catalog.}
\label{examples_sf}
\end{figure*}

\subsection{Enhanced AGN activity in large-scale structures} 

The enhancement of galaxies hosting AGN, in large-scale structures
reported in \citet{gi03} and substantiated further in this
investigation, appears to indicate the physical scale most nuturing
for accretion onto SMBHs.  The fraction of galaxies hosting AGN
(Fig.~\ref{agn_fraction}b) is significantly enhanced within a redshift
interval ($0.63 \leq z \leq 0.76$) dominated by two prominent redshift
spikes ($z=0.67$ and $z=0.73$) that are overpopulated with both AGN
\citep{gi03} and galaxies \citep{ci02,ad05}.  The dimensions of these
structures are estimated to be $\sim$10 Mpc in transverse
extent\footnote{Silverman et al. (2007) demonstrate that both
structures do extend beyond the central 1 Ms region and provide new
measures of their angular size.}  and $\sim37$ ($z=0.67$) and $\sim28$
($z=0.73$) Mpc in depth (R. Gilli, private communication).  These two
structures have been shown \citep{gi03,ad05} to be varied in their
dynamical state, though both have signs of ongoing collapse
\citep{ad05}.  The 'wall' or 'sheet' at $z=0.67$ is described as a
loose structure, not yet virialized, and with an embedded compact
feature, possibly an infalling group.  The $z=0.73$ 'wall' is more
compact with a central dense core due to the presence of a cluster
\citep{ci02,gi03}, and appears to have significant surrounding
substructure \citep{ad05} that may eventually collapse into a massive
Virgo-type cluster in the local universe.  The difference in galaxy
populations within these two structures appears to reflect their
dynamic state following the well-known SFR-density relation: galaxies
in the $z=0.67$ structure (Fig.~\ref{color_mag}b) are mainly blue
while the $z=0.73$ structure is dominated by red (i.e., evolved)
galaxies.

Most remarkably, the colors of the host galaxies of these
moderate-luminosity AGN\footnote{It is worth noting that \citet{ge07}
recently reported a color-dependency for AGN residing in denser
environments with rest-frame colors ($U-B\sim0.8$) coincident with
blue galaxies.}, within the redshift interval $z=$~0.63--0.76,
preferentially fall in the ``green valley'' (Fig.~\ref{color_histo}b).
The color distribution is asymmetric, with many hosts closer to the
red sequence than the blue galaxy peak.  As evident in
Figure~\ref{color_mag}b, this color profile is composed of a broad
distribution of galaxies residing in the $z=0.67$ redshift spike and a
more compact, red ($U-V\sim0.8$) group in the $z=0.73$ redshift spike.
It appears that the AGN follow a SFR-density type relation similar to
the galaxies.  In Figure~\ref{examples_valley}, we show example
optical spectra and $HST$ images of four AGN that have $z=$~0.63--76
and colors placing them in the ``green valley''.  All have no evidence
for an embedded AGN (e.g., H$\beta$ or Mg II emission lines) and a
relatively mild 4000~\AA~break.  The spectrum of source \#85 is
characteristic of an E+A galaxy \citep[i.e., poststarburst;
][]{dr83,bl04} based on the absorption-line strength (EW=6.7 ${\rm
\AA}$) of H$\delta$ and weak [OII] emission.  E+A galaxies are thought
to have undergone a starburst phase $\sim1$ Gyr ago based on the
strong contribution from A stars and lack of O and B stars indicative
of very recent ($\sim0.01$ Gyr) or ongoing star formation.  This
object is only one of fifteen, over the full redshift range, that fall
within the ``green valley'' and has a spectrum characteristic of an
E+A galaxy.  Therefore, most of these galaxies have not had a major
starburst episode within 1--2 Gyr in their past.  Most spectra seem to
be characteristic of a galaxy, with gradual ongoing star formation,
which is best fit by the Type 2 and 3 average galaxy templates shown
in Figure 3 of \citet{wo03}, effectively separating blue and red
galaxies.  A larger sample is required to measure statistically the
fraction of AGN hosts with poststarburst signatures that have been
shown to be common in AGN from the SDSS \citep{vanden06}.  {\it We
conclude that within this narrow redshift interval, containing two
overdense structures, the enhancement of AGN activity within the ``green
valley'' signifies an important link between the evolution of SMBHs
and their host galaxies though not overwhelmingly in poststarburst
systems.}

\begin{figure*}

\includegraphics[scale=0.60,angle=0]{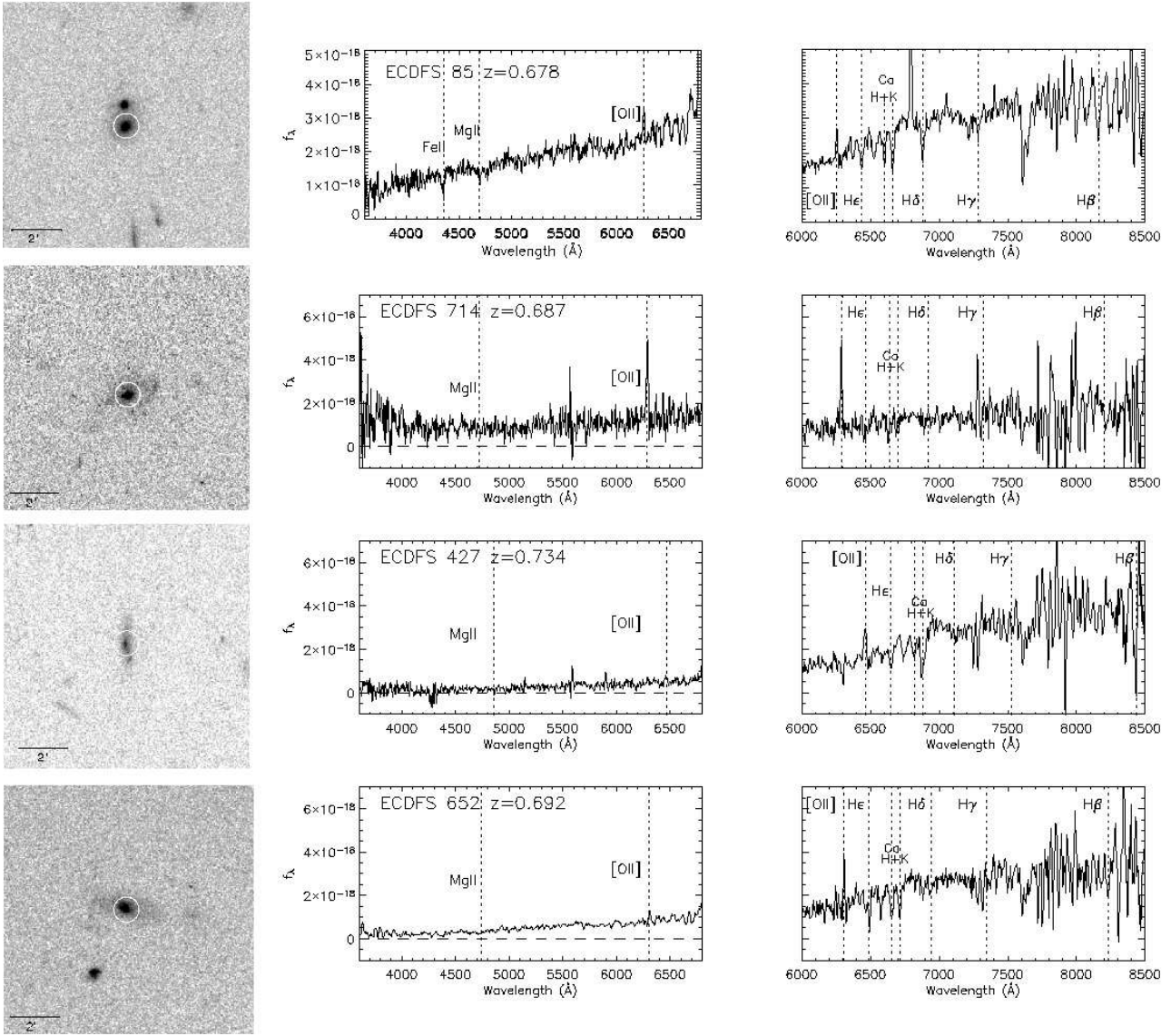}
\caption{Examples of AGN hosts with $0.5<U-V<1.0$ and spectroscopic
redshifts placing them in the vicinity of the redshift spikes
($z\sim0.7$).  The {\it HST}/ACS $V_{606}$-band images are shown with a log
scaling. Optical spectra are described in the caption of
Figure~\ref{examples_sf}.  The source numbers follow those presented in the
\citet{le05} X-ray catalog.}
\label{examples_valley}
\end{figure*}

\subsection{Optical morphology of AGN hosts}

\label{morphology}

We highlight the fact that AGN hosts show signs of a bimodal
distribution in their colors but not in their morphological properties
\citep{bo06}.  To illustrate, we have marked those AGN in
Figure~\ref{color_mag}a that have a radial surface brightness profile
(i.e., S\'{e}rsic index, $n$), provided by \citet{ha07},
characteristic of a bulge-dominated galaxy ($2.5<n<8$).  The
S\'{e}rsic profile of a typical disk-dominated (e.g., late-type)
galaxy is that of an exponential function ($n=1$) while that of a
bulge-dominated galaxy (e.g., early-type) is the $r^{1/4}$-law (de
Vaucoulers profile; $n=4$).  We find that there are many
bulge-dominated galaxies with bluer colors (i.e., ongoing star
formation) than a typical red-sequence galaxy and there are a few
disk-dominated galaxies that are red.  The {\it HST}/ACS images of the
red disk-dominated galaxies hosting AGN show that most (5 out of 7
with $U-V>1.1$) of them are highly-inclined or edge-on spirals that
probably have significant dust extinction.  To investigate further, we
plot in Figure~\ref{morph_color}, the S\'{e}rsic index versus
rest-frame color ($U-V$) for our luminous ($M_V<-20.7$) galaxies and
AGN.  We further require here that the science flag provided by
\citet{ha07} is 1 for all galaxies that guarantees that GALFIT
\citep{pe02} converged; this condition is met for 71 of the 85
luminous host galaxies with morphological parameters given in
\citet{ha07}.  The rest-frame color distribution of this cleaned
sample is equivalent to that used in previous analyses since the
removed hosts span the full range of color.  Symbol types for the AGN
correspond to the three redshift intervals used throughout this work.
First, it is evident that the requirement for AGN host galaxies to be
bulge-dominated \citep[e.g.,][]{ka03} continues up to $z\sim1$
\citep{gr05} since 75\% have $n>2.5$ and the fraction\footnote{The
method to determine the AGN fraction (Table~\ref{table_fraction}; Index
8-12), using our cleaned sample (science flag=1), is described in
$\S$~\ref{text:agn_fraction} with $n$ replacing $U-V$.} of galaxies
hosting AGN as a function of $n$ sharply rises above this value
(Fig.~\ref{agn_fraction_morph}a; Table~\ref{table_fraction}).  Second,
AGN hosts preferentially become bluer with redshift but retain their
bulge-dominated morphology.  We conclude that the blue colors of AGN
hosts up to $z\sim1$ are not indicative of star formation in
disk-dominated systems.

\begin{figure}
\includegraphics[scale=0.50,angle=90]{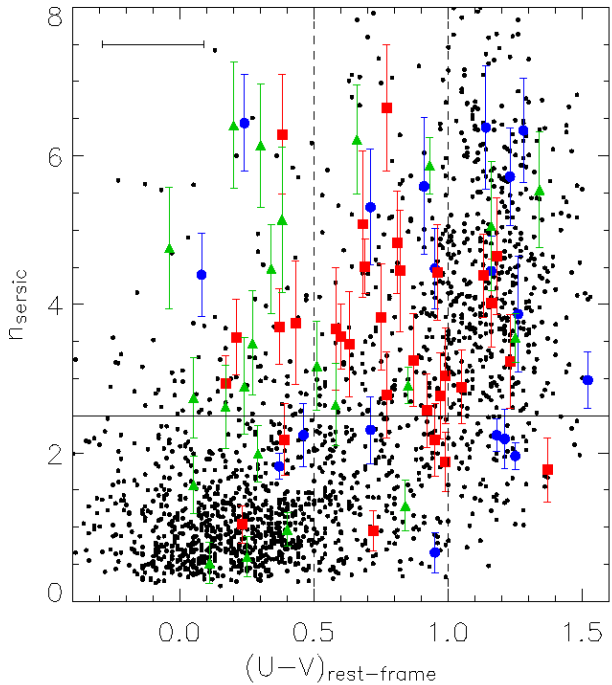}
\caption{Morphology--color relation of luminous ($M_V<-20.7$) galaxies
with (large colored symbols) and without (small black dots) AGN.  The
large symbols denote the following redshift ranges: $z =$~0.4--0.63
(blue circles), $z =$~0.63--0.76 (red boxes), $z =$~0.76--1.1 (green
triangles).  Errors (1$\sigma$) associated with $n$ are those reported
in \citet{ha07}.  The bar in the upper left corner is the mean
$\pm1\sigma$ error on $U-V$ for the AGN host galaxies.  The horizontal
line divides those galaxies that have either a disk-dominated
($n<2.5$) or bulge-dominated ($n>2.5$) morphology as defined in
\citet{bl03}.  The vertical lines highlight the ``green valley''.}
\label{morph_color}
\end{figure}

It is clear that the fraction of bulge-dominated galaxies hosting AGN
is higher for those with blue colors ($U-V<0.7$); this is shown in the
top left quadrant of Figure~\ref{morph_color}.  We measure the AGN
fraction (see Table~\ref{table_fraction}), as detailed in
$\S$~\ref{text:agn_fraction}, and find that $21.3\pm5.0\%$ of the
galaxies with $n>2.5$ and $U-V<0.7$ harbor moderate-luminosity AGN.
This is over three times greater than that found for the entire sample
irrespective of color or morphology ($6.4\pm0.8$\%;
$\S$~\ref{text:agn_fraction}).  In contrast, the AGN fraction of
purely disk-dominated galaxies ($n<1.5$) is much lower
($1.0\pm0.4$\%).  We conclude that blue, bulge-dominated galaxies have
a relatively-high AGN fraction, possibly due to a combination of their
massive bulges and gas content.

\begin{figure}
\includegraphics[scale=0.65,angle=90]{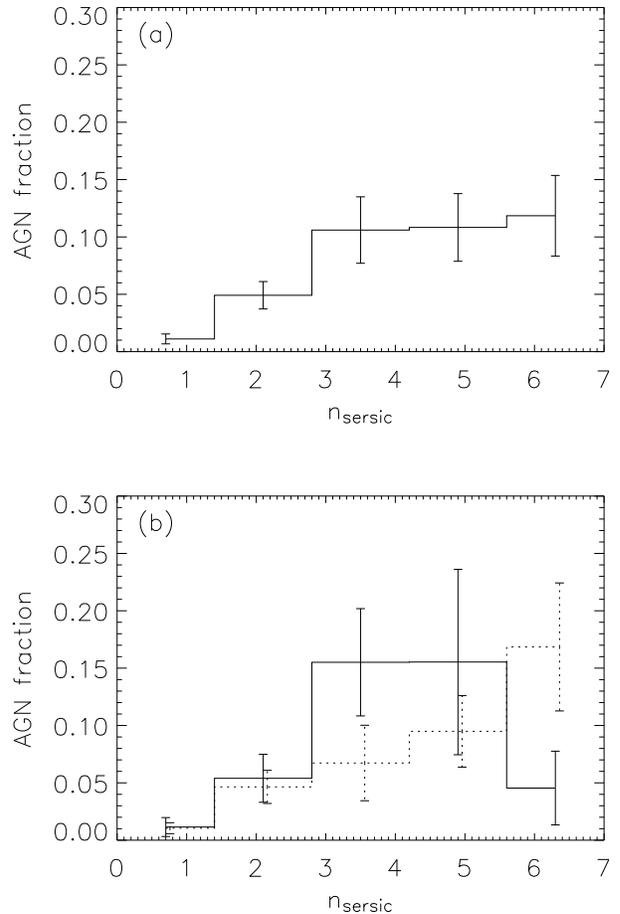}
\caption{Fraction of galaxies ($M_V<-20.7$) hosting AGN as a function
of their morphology (i.e., S\'{e}rsic index $n$) for the entire sample
(a).  In panel $b$, we have split the sample into those in (solid) and
out (dashed) of the redshift interval $0.63 \leq z \leq 0.76$ as done
in Figure~\ref{agn_fraction}.  AGN clearly reside in bulge-dominated
(Panel $a$; $n\gtrsim3$) galaxies with those in large-scale structures
(Panel $b$; solid line) possibly having a higher fraction with faint
disks that effectively softens their S\'{e}rsic index ($n\sim3$).}
\label{agn_fraction_morph}
\end{figure}

We further illustrate the color-morphology of our AGN host galaxies by
displaying in Figure~\ref{bulges} {\it HST}/ACS color images of six
AGN host galaxies in the GOODS area with $n>2.5$ sorted by rest-frame
$U-V$ color.  Clearly, the most prominent feature is the bright bulge
for all host galaxies.  The bluest host galaxy J033213.2$-$274241,
seems to be undergoing a merger.  It appears that faint disks may be
prevalent in AGN residing within the ``green valley''; this is evident
in Figure~\ref{examples_valley} as well.  Star-forming substructures
may be present in the form of arcs or spiral arms in a few host
galaxies (J033226.8$-$274145, J033246.4$-$275414).  For comparison,
the bottom right panel shows a typical red-sequence elliptical galaxy
(\hbox{J033209.7$-$274248}).  In Figure~\ref{disks}, we display color
images of disk-dominated ($n<2.5$) AGN host galaxies.  Bright,
star-forming regions are ubiquitous in the form of knots that appear
to lie within spiral arms in most examples.

\begin{figure*}
\includegraphics[scale=1.0,angle=0]{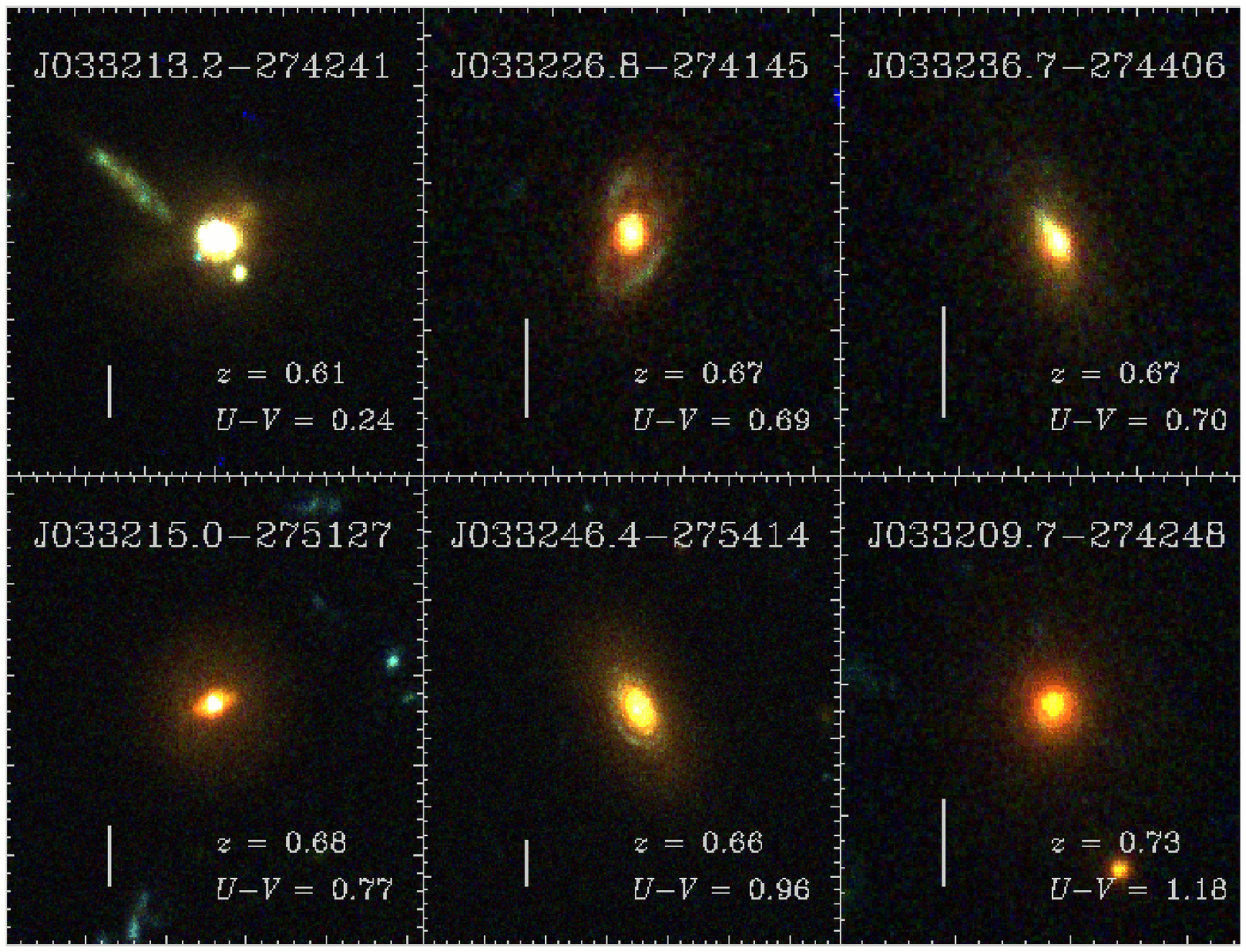}
\caption{Color {\it HST}/ACS postage-stamp images of bulge-dominated
($n>2.5$) host galaxies located in the GOODS region.  Colors correspond to ACS
$B_{435}$ (blue), $V_{606}$ (green), and $z_{850}$ (red) bandpass images.  In
each postage-stamp image, we indicate the source name (top), the redshift and
rest-frame $U-V$ color of the galaxy (lower right), and a vertical line of
length $1.5\arcsec$ for scaling reference.  The images have been sorted by
rest-frame $U-V$ color such that the bluest source is located in the upper left
panel and the reddest source is shown in the lower right panel.  
}
\label{bulges}
\end{figure*}

\begin{figure*}
\includegraphics[scale=0.85,angle=0]{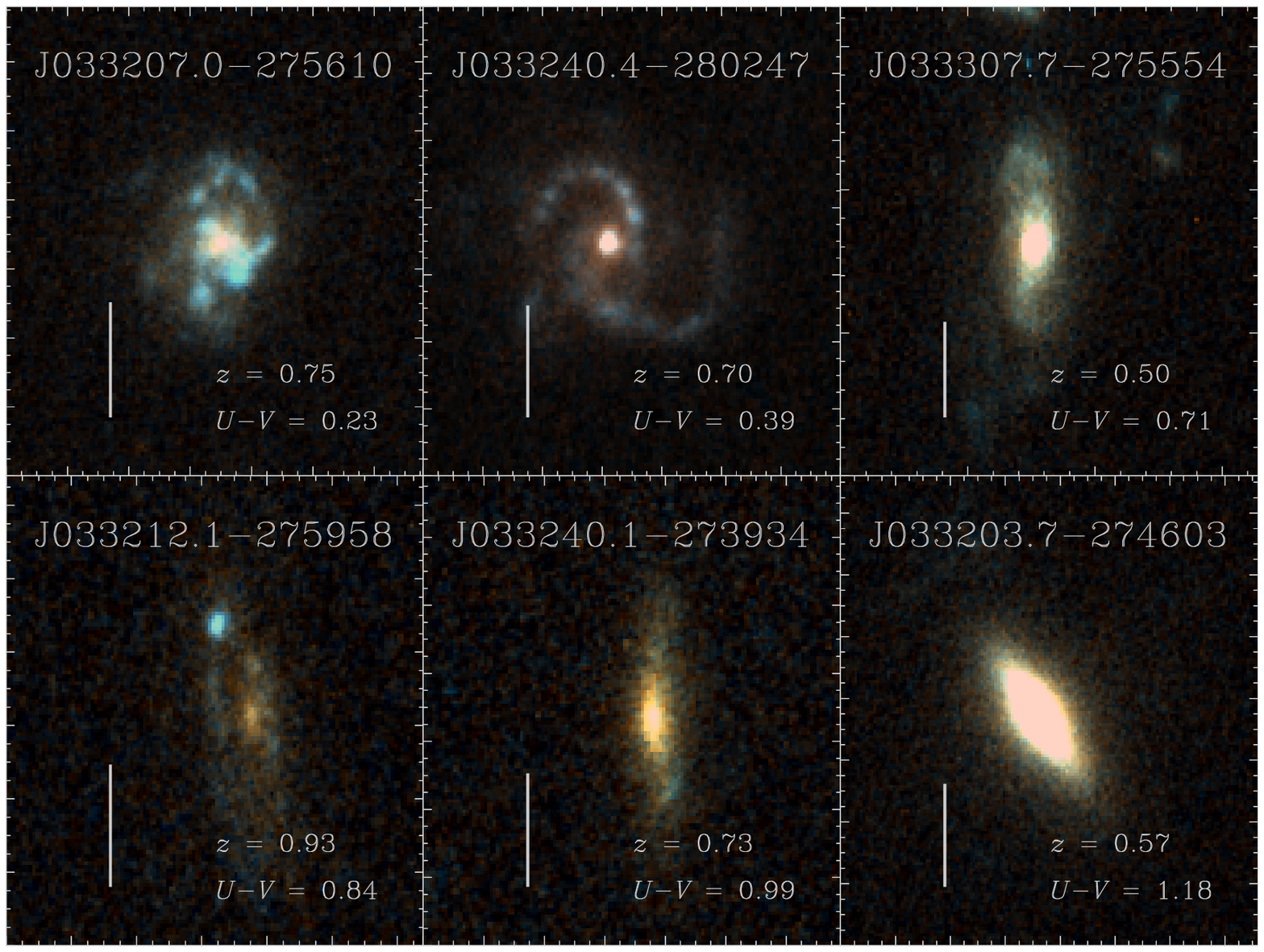}
\caption{Color {\it HST}/ACS postage-stamp images of disk-dominated
($n<2.5$) host galaxies located in the GEMS area.  Colors correspond
to ACS $V_{606}$ (blue), ($V_{606}$+$z_{850}$)/2 (green) and $z_{850}$
(red) bandpass images.  The labels are described in Figure~\ref{bulges}.}
\label{disks}
\end{figure*}

We remark that a more thorough analysis of the morphological
structures of these host galaxies and their relation to the general
galaxy population is required and beyond the scope of this paper.  We
highlight the following complications that are relevant for
interpreting these results and will be taken into consideration in a
future analysis: (1) S\'{e}rsic fits using de Vaucoulers profiles have
been shown to have significant uncertainties \citep[e.g.,][]{ha07},
(2) the presence of an embedded, optically-faint AGN is likely to
affect the measure of structural parameters, such as $n$, (3) accurate
discrimination between bulge and disk-dominated galaxies is difficult
since S\'{e}rsic indices are known to have significant overlap for
these populations \citep[e.g.,][]{sa06}, and (4) one must consider
redshift-dependent surface brightness effects.  A fair number of AGN
host galaxies have intermediate S\'{e}rsic indices ($1.5<n<3$;
Figure~\ref{morph_color}) and evidence of faint disks as described
above.  A careful consideration of these issue will enable us to
assess the dependence of host-galaxy morphology with environment
(Figure~\ref{agn_fraction_morph}b}).  There is suggestive evidence
that S\'{e}rsic indices of hosts within large-scale structures
(Figure~\ref{morph_color}; red squares) are shifted to lower values
($n\sim3$).  The frequency of bulge-dominated disk galaxies among the
hosts of AGN may be an important test of merger-based evolution
models.

\subsection{Are AGN driving the evolution of the general galaxy population?}
\label{text:mergers}

Numerical simulations \citep{sp05b} demonstrate that accreting SMBHs
may be a necessary ingredient in the formation of massive elliptical
galaxies from the mergers of gas-rich spiral galaxies.  AGN feedback
can potentially suppress star formation \citep{di05} and drive (i.e.,
accelerate) galaxies onto the red sequence \citep{cr06,hop06a}.  The
short timescales ($<$ 1 Gyr) for an AGN to provide significant
feedback to expel the gas may then instill the color bimodality of
galaxies \citep{sp05a} since galaxies with intermediate colors are
thought to be evolving rapidly.  Strong observational evidence
supporting this scenario has been elusive.  We now explore whether our
AGN sample with well determined host colors and luminosities offers
further insight into the AGN-galaxy connection.

The color distribution of AGN host galaxies in our study further
substantiates the aformentioned model of galaxy evolution.  The
fraction of galaxies harboring AGN is significantly enhanced
(Fig.~\ref{agn_fraction}) in the ``green valley'', a region on the
color-magnitude diagram where galaxies are thought to migrate from
blue to red.  This higher incidence is associated with galaxies
residing in a redshift interval $0.63 \leq z \leq 0.76$ dominated by
two large-scale structures that are overdense and have an ongoing assembly
of substructure (i.e., galaxy groups and clusters).  Within this same
redshift interval, the color distribution of the underlying galaxy
population (Fig.~\ref{galaxy_distributions}) has rapidly evolved as
shown by (1) a significant reduction in the numbers of blue galaxies,
a dominant population at higher redshifts ($z>0.76$), (2) a redward
shift of the blue peak, and (3) a dramatic emergence of a red sequence
compared to previous epochs ($0.76<z \leq 1.1$; dotted line).  This
narrow redshift interval ($0.63 \leq z \leq0.76$), with an elapsed
time of 0.73 Gyr, offers a compressed window of the more passive
galaxy evolution that occurs over longer timescales of $\sim$5 Gyr
($0.4<z<1.1$) as illustrated in Figure~\ref{color_mag}b.

\begin{figure}

\includegraphics[scale=0.55,angle=0]{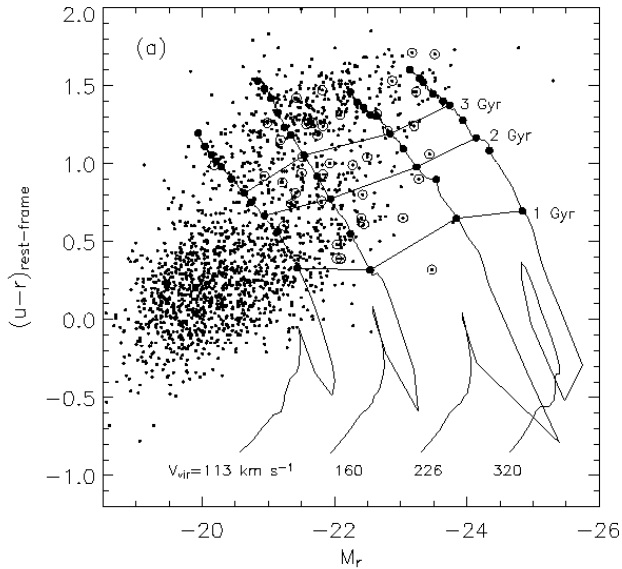}
\includegraphics[scale=0.65,angle=0]{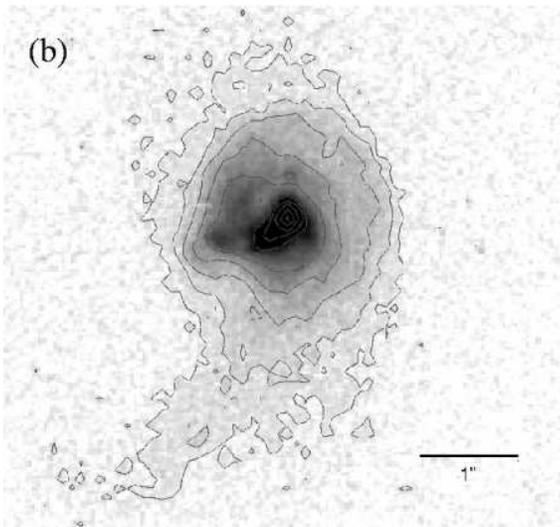}
\caption{(a) Color-magnitude relation of galaxies ($0.63 \leq z \leq 0.76$)
associated with large-scale structures and AGN-influenced model
evolutionary tracks from \citet{sp05a}.  Galaxies hosting AGN are
further marked by an open circle.  Theoretical data for merging
galaxies is overplotted for galaxies with virial velocities as
shown. The large filled circles start from 1 Gyr after the inital
starburst phase and each spacing corresponds to 0.5 Gyr. (b) $HST$
$V_{606}$-band image of source \#66 ($z_{phot}=0.69$;
$L_{\rm X}=1.5\times10^{42}$ erg s$^{-1}$) that is the only AGN host galaxy
shown above ($M_r=-23.47; u-r=0.32$; $n=3.56\pm0.018$) with a
possible merger timescale less than 1 Gyr.}
\label{color_tracks}
\end{figure}

In addition, numerical simulations demonstrate that the color
evolution of a merger event slows upon approach to the red sequence
\citep{sp05a}.  In Figure~\ref{color_tracks}, we show the
color-magnitude relation for galaxies and AGN within the redshift
interval $0.63\leq z \leq 0.76$ with the evolutionary tracks of
mergers, including SMBH feedback, with virial velocity $v_{\rm
vir}$=113, 160, 226, and 320 km s$^{-1}$ overplotted \citep{sp05a}.
The fact that these models do not account for dust reddening should
not severely impact our subsequent findings.  Here, we use SDSS
photometric bands ($u$, $r$) to utilize these models and have
converted the photometry to the Vega system.  To account roughly for
redshift evolution from $z\sim0.7$ to the present, we have shifted the
model magnitude $M_r$ by $-1$ based on values given for $M^{\star}_B$
in the literature \citep[see Table 5 of ][]{fa06}.  The evolutionary
tracks have ages up to $\sim$5.5 Gyr\footnote{It is just a coincidence
and most likely of no physical significance that these large-scale
structures in the CDF-S are present at $z\sim0.7$, a redshift singled out
by \citet{sp05a}, as a possible formation epoch of elliptical galaxies
based an elapsed time of 5.5 Gyr to complete a merger sequence.} with
the first data point corresponding to 1 Gyr after the initial
starburst phase.  Based on these model curves, over 1 Gyr has elapsed
since the starburst phase concluded for almost all AGN hosts.  The
majority of AGN hosts cover a broad range of age between 1--4 Gyr and
virial velocity between 113 and 226 km s$^{-1}$.  As shown in
Figure~\ref{color_tracks}, a large fraction of the merger sequence
($t\approx$~2--4 Gyr) is spent on approach to the red sequence that
should impart an asymmetry in the color distribution of the hosts of
AGN.  In Figure~\ref{color_histo}b, there may be signs of this since
the distribution is skewed with many hosts falling near the blue edge
of the red sequence, although a larger sample is required to a make a
definitive claim.  This effect coupled with our sample size may
explain the small numbers of host galaxies with clear poststarburst
($t \lesssim$~1--2 Gyr) signatures in their optical spectra
(Fig.~\ref{examples_valley}).  We further speculate that the AGN
hosts, situated at the entrance to the red sequence, may either redden
with age or further merge with galaxies already on the red sequence
\citep[i.e., `dry mergers';][]{vd05,be06} in rich environments that
can effectively move them to redder colors and high
luminosities/masses since there are few massive ($v_{\rm vir}>226$ km
s$^{-1}$) and luminous progenitors that could populate the luminous
end ($M_{r} \lesssim -22.5$) of the red sequence.  We conclude that
the color distribution of AGN hosts further substantiates a
coevolution scenario due to mergers and interactions that are
effectively nurtured in $\sim10$ Mpc scale structures.

The role of major mergers in triggering AGN has met observational
scrutiny even up to $z\sim1$, where the merger rate is expected to be
higher.  Morphological studies \citep{gr05,pi07} have yet to find
X-ray selected AGN at $z>0.5$ in such disturbed systems.  Here, we
confirm the findings of \citet{na07} that hard (i.e., obscured) AGN do
not populate a distinct region of the color-magnitude diagram as
predicted by numerical simulations of merging galaxies that trigger a
pre-quasar (optically hidden) phase \citep{hop05}.  We have further
marked those AGN in Figure~\ref{color_mag} with hardness ratios
\hbox{$HR=(H-S)/(H+S)$} indicative of X-ray absorption ($HR>-0.2$;
$\Gamma=1.9$ with \hbox{$N_{\rm H}=10^{22}$~cm$^{-2}$} at $z=0$).  The
hardness ratio is a measure of the relative numbers of observed X-ray
counts in the soft ($S$; 0.5--2.0 keV) and hard ($H$; 2--8 keV) energy
bands.  We see that hard sources are present over a wide range of
rest-frame colors.  This is not unexpected since AGN are known to
usually have parsec-scale molecular tori \citep{an93} with substantial
absorbing columns that bear no relation to the presence of star
formation.  The lack of mergers may reflect the limited sample used to
date, since a pair of galaxies undergoing a major merger at $z\sim0.7$
may only be resolved by {\it HST} for less than $\sim$1 Gyr of the
merger cycle \citep[see Fig.~11 of ][]{sp05b}.  As is evident in
Figure~\ref{color_tracks}, the galaxy samples in the \hbox{E-CDF-S}
are severely limited for the region on the color-magnitude diagram
likely to have merger events between massive ($v_{\rm vir}>160$ km
s$^{-1}$) galaxies on a timescale of $\sim$1 Gyr after the initial
starburst has ended.  Three hosts out of our entire sample of 109 AGN
are located on the color-magnitude diagram in a region with $v_{\rm
vir}>160$ km s$^{-1}$ and $\tau < 1$ Gyr which may be indicative of a
major merger.  As shown in Figure~\ref{color_tracks}a, one (\#66) of
them falls within our narrow redshift interval $0.63 \leq z \leq 0.67$
with $M_r=-23.47$ and $u-r=0.32$.  In Figure~\ref{color_tracks}$b$, we
show the $HST$ $V_{606}$-band image that exemplifies a complex
morphology in the nuclear region and tidal features on scales of
$\sim$10 kpc (1\arcsec~=~7.1 kpc at $z=0.69$) that most likely arise
from a major merger of massive galaxies.  A second example
(J033213.2$-$274241), shown in the top left panel of
Figure~\ref{bulges}, also shows signs of interaction.  We surmise that
larger area {\it HST} surveys of deep extragalactic fields, such as
COSMOS \citep{sc06} with an area coverage of 1.8 deg$^2$ (5.4~$\times$
the area of the \hbox{E-CDF-S}) will provide improved statistics to
assess adequately the role of mergers in triggering SMBH accretion.

Finally, we note an alternative to the merger scenario in which
galaxies may be more favorable to AGN activity simply due to the
presence of a massive bulge and disk that provide the two required
ingredients \citep[i.e., a SMBH and a reservoir of gas for
accretion;][]{ka06}.  This is possible since a high fraction
($\sim21\%$) of blue spheroids ($n>2.5$; $U-V<0.7$) host
moderate-luminosity AGN and there is a slight enhancement of AGN
activity in the `field' for galaxies in the ``green valley''
(Fig.~\ref{agn_fraction}b; dotted line).  HST images of host galaxies
residing within the ``green valley'' (Fig.~\ref{examples_valley})
appear to have bulges and faint disks.  If this were the case,
however, we would then not expect such a strong increase in the AGN
fraction as a function of environment.

\section{Summary}

We identified a sample of 109 X-ray selected AGN in the \hbox{E-CDF-S}
with moderate luminosities ($41.9\leq \log~L_{0.5-8.0~{\rm keV}}\leq
43.7$) to investigate the rest-frame colors of their host galaxies.
These AGN have been selected from a parent sample of 5,549 galaxies
from COMBO-17 and GEMS with $0.4 \leq z \leq 1.1$.  Optical spectra
are available for 48\% of the sample; these provide assurance of the
accuracy of the photometric redshifts and show that no strong AGN
signatures are present, confirming the stellar nature of their
rest-frame colors.

We find that the broad distribution of host-galaxy colors of
moderate-luminosity AGN is due to both (1) the strong color evolution
of the underlying luminous ($M_V<-20.7$) and bulge-dominated ($n>2.5$)
galaxy population and (2) an enhancement of AGN activity in
large-scale structures.  We draw the following three main points:

\begin{itemize}

\item The host galaxies of X-ray selected AGN have a color bimodality
when excluding a redshift interval $0.63 \leq z \leq 0.76$, which contains two
redshift spikes at $z=0.67$ and $z=0.73$.  Galaxies hosting AGN at $z
\lesssim 0.6$ are preferentially red with many falling along the red
sequence.  At $z\ga0.8$, a distinct, blue population of host galaxies
is prevalent with colors similar to the star-forming galaxies.

\item The fraction of galaxies hosting AGN has a prominent peak in the
``green valley'' that is primarily due to enhanced AGN activity in
large-scale structures.  Within the redshift interval $0.63 \leq z
\leq 0.76$, the AGN fraction reaches a peak value of $\sim15\%$ at
$U-V\sim0.8$.  Over the color interval $0.5<U-V<1.0$ we find the AGN
fraction to be $12.8\pm2.9\%$, significantly ($2.6\sigma$) higher than
that measured in the `field' (i.e., over all other redshifts;
$7.8\pm2.5\%$).

\item We find that AGN continue to preferentially reside in luminous
bulges up to $z\sim1$.  A large fraction (75\%) of AGN host galaxies
with $M_V<-20.7$ have S\'{e}rsic indices $n>2.5$, even those at
$z\sim1$ that primarily have blue rest-frame colors.  Blue,
bulge-dominated galaxies ($n>2.5$, $U-V<0.7$) are most hospitable for
AGN activity based on a measured AGN fraction of $21.3\pm5.0\%$ (four
times the fraction of the `field' sample).

\end{itemize}

The overabundance of AGN associated with the redshift spikes found in
the \hbox{E-CDF-S} and their special location in the color-magnitude
relation highlight the importance of environment, on large scales
($\sim10$ Mpc), to influence the evolution of AGN and their host
galaxies.  The richness of these structures (i.e., galaxy overdensities,
group/clusters in early stages of formation) alludes to mergers as a
dominant mechanism to trigger AGN activity, quench star formation and
to drive subsequent migration of galaxies from the blue cloud to the
red sequence.  We compare the color-magnitude relation of our AGN host
galaxies with evolutionary tracks of merging galaxies from
\citet{sp05a} that incorporate AGN feedback.  Our AGN host galaxies
have colors and morphologies (i.e., bulge-dominated) indicative of
evolved systems that had undergone a starburst phase $\approx$ 1--4
Gyr, and a possible major merger before being observed.  These
timescales qualitatively agree with optical spectra that do not show
overwhelming starburst signatures.  These timescales also naturally
explain why so few X-ray selected AGN host galaxies appear to be
undergoing major mergers.  Larger area $HST$ imaging surveys such as
COSMOS will statistically sample these rare events with timescales
less than 1 Gyr.

The \hbox{E-CDF-S} is a unique survey field with a fortunate alignment
of large-scale structures for such studies.  Our findings exemplify
the complexities that must be disentangled to determine underlying
relationships between AGN and their host galaxies.  Much optical and
near-infrared followup is forthcoming in the \hbox{E-CDF-S} that will
further our understanding of the connection between coevolution of AGN
and galaxies.

\acknowledgments

We are especially grateful to the referee for providing insightful
comments that strengthened the overall content of this work.  We also
thank L.~Guzzo, K. Iwasawa, A. Merloni, K. Nandra, and I. Strateva for
helpful discussions and suggestions.  We also recognize the
contribution of the GEMS group that provided an early version of their
catalog, V. Springel for supplying the model galaxy tracks, and
C.~Wolf for computing updated photometry.  Support for this work was
provided by NASA through $Chandra$ Award Number G04-5157A (BDL, WNB,
DPS).  RG, CV and PT acknowledge partial support by the Italian Space
agency under the contract ASI--INAF I/023/05/0.

"Some of the data presented in this paper were obtained from the
Multimission Archive at the Space Telescope Science Institute
(MAST). STScI is operated by the Association of Universities for
Research in Astronomy, Inc., under NASA contract NAS5-26555. Support
for MAST for non-{\it HST} data is provided by the NASA Office of Space
Science via grant NAG5-7584 and by other grants and contracts."



Facilities: \facility{VLT(VIMOS)}.

\clearpage



\end{document}